\PassOptionsToPackage{numbers,sort&compress}{natbib}
\documentclass[preprint,10pt, a4paper]{elsarticle}

\usepackage[margin=2.5 cm]{geometry} 
\usepackage{graphicx}
\usepackage{array}[=2016-10-06]
\usepackage{dcolumn}
\usepackage{bm}
\usepackage{lipsum}
\usepackage{latexsym}
\usepackage{amsfonts}
\usepackage{amssymb}
\usepackage{amsmath}
\usepackage{bbold}
\usepackage{booktabs}
\usepackage{float}
\usepackage{verbatim}
\usepackage{overpic}
\usepackage{siunitx}
\usepackage{physics}
\usepackage{rotating}
\usepackage[utf8]{inputenc} 
\usepackage{xspace}

\usepackage{listings}
\usepackage[colorlinks=true, linkcolor=blue, citecolor=blue, urlcolor=blue]{hyperref}
\usepackage{xcolor}
\usepackage{pdflscape}
\usepackage{tabularx}
\usepackage[referable]{threeparttablex}
\newcommand{\mrn}{\mathrm{n}}
\newcommand{\mrp}{\mathrm{p}}
\newcommand{\mrs}{\mathrm{s}}
\newcommand{\mre}{\mathrm{e}}

\newcommand{\cmax}{c_{k,\text{max}}}

\DeclareMathOperator{\arcsinh}{arcsinh}

\newcommand{\vect}[1]{\bm{#1}}
\newcommand{\matr}[1]{\mathbf{#1}}

\usepackage{calligra}
\DeclareMathAlphabet{\mathcalligra}{T1}{calligra}{m}{n}
\DeclareFontShape{T1}{calligra}{m}{n}{<->s*[2.2]callig15}{}
\newcommand{\scriptr}{\mathcalligra{r}\,}

\usepackage{circuitikz}
\usepackage[nameinlink,capitalise]{cleveref}

\begin{document}

\date{\today}
\title{Surrogate-accelerated parameterisation of physics-based Li-ion battery models}
\author[1]{A. Emir G\"umr\"uk\c{c}\"uo\u{g}lu\corref{cor1}}
\ead{emir.gumrukcuoglu@port.ac.uk}
\author[1]{Josh Pearson}
\author[1,2]{Jamie M. Foster}
\author[1]{James Burridge }

\affiliation[1]{organization={School of Computing, Mathematics \& Physics,
University of Portsmouth},
addressline={Lion Terrace},
city={Portsmouth},
 citysep={}, 
postcode={PO1 3HF},
country={UK}}

\affiliation[2]{organization={The Faraday Institution},
addressline= {Quad One, Becquerel Avenue, Harwell Campus},
city={Didcot},
citysep={},
postcode={OX11 0RA},
country={UK}}

\cortext[cor1]{Corresponding author}

\begin{abstract}
Physics-based lithium-ion battery models provide access to physically meaningful internal electrochemical states and processes, but cell-specific parameter inference from terminal current--voltage data is computationally expensive and limited by identifiability.
We present a surrogate-accelerated inverse framework based on a single particle model with electrolyte dynamics (SPMe).
Its forward map uses our Artiphy surrogate framework for rapid, differentiable evaluation of voltage and selected internal states.
After rescaling to remove exact structural redundancies, we infer non-redundant transport, kinetic and capacity parameter groups, including concentration-dependent solid and electrolyte diffusivities.
Synthetic voltage data from a Doyle-Fuller-Newman (DFN) model under a drive-cycle protocol provide a benchmark with known reference parameters and controlled model discrepancy.
The inferred SPMe reproduces the benchmark voltage with an error of order $1~\unit{\milli\volt}$ and recovers electrode capacities well. Positive-electrode diffusivity is recovered accurately over much of the probed stoichiometric range.
Local sensitivity and Fisher information analysis identifies correlated kinetic--Ohmic and electrolyte transport directions, and shows how localised information and the global diffusivity parameterisation can restrict variation in the reconstructed diffusivity across stoichiometries
despite weak voltage sensitivity to negative-electrode diffusion over much of the drive cycle.
These results represent a step towards rapid physics-based in-silico parameterisation and reduced reliance on destructive cell characterisation.
\end{abstract}

\maketitle
\newpage
\tableofcontents
\newpage
\section{Introduction}

Physics-based continuum models of lithium-ion batteries, notably the Doyle--Fuller--Newman (DFN) model \cite{doyle93,1994FuDoNe} and reduced counterparts such as the single particle model (SPM) and single particle model with electrolyte dynamics (SPMe) \cite{bro22}, are valuable because their parameters and internal states have electrochemical meaning.
Unlike conventional equivalent-circuit models, which describe terminal behaviour through effective circuit elements, these models can expose internal electrochemical quantities that are not directly
measurable in routine testing, including electrode stoichiometries, electrolyte concentration gradients, reaction overpotentials and transport limitations.
This interpretability motivates their use in battery management, cell characterisation, lifetime modelling, and degradation-risk assessment \cite{bro22,LIU2022105176,Guo2024}.

However, these models are not automatically predictive for a particular cell. The DFN model is already an effective continuum description of a heterogeneous multiscale system \cite{bro22, LIU2022105176}. In particular, the pseudo-two-dimensional DFN geometry replaces the real porous electrode microstructure by a through-cell coordinate coupled to representative spherical particles, while quantities such as tortuosity, effective transport coefficients and particle radii summarise microstructural information that is not explicitly resolved \cite{Korneev2020,Kirk2022}.

Cell-specific prediction requires these effective parameters to be determined for the cell of interest. This generally requires several complementary experimental techniques, including cell teardown and imaging, half-cell open-circuit potential (OCP) measurements, galvanostatic or potentiostatic intermittent titration, and electrochemical impedance spectroscopy: no single experiment determines the full set of geometrical, thermodynamic, kinetic and transport quantities \cite{Wang_2022}. Comprehensive workflows have been demonstrated for commercial cells, but they are experimentally demanding, commonly involve destructive teardown and half-cell reconstruction, and may still require selected parameters to be fitted or adjusted against full-cell data \cite{ecker15, chen20, zulke21}. Dynamic current--voltage data therefore offer a complementary, non-invasive route to cell-specific parameterisation \cite{Namor2017}.

Parameter estimation for physics-based lithium-ion battery models from applied-current and terminal-voltage data has substantial precedent, spanning SPM, SPMe and full DFN formulations and a wide range of optimisation and Bayesian inference methods \cite{Santhanagopalan2007,FORMAN2012263,ait20,MIGUEL2021103388,ANDERSSON2022230859}. The resulting inverse problems differ considerably in model fidelity, excitation protocol, measurement modality, and the number and type of fitted parameters \cite{MIGUEL2021103388,ANDERSSON2022230859}.
Many studies reduce the problem through sensitivity-based parameter selection \cite{Jin2018,Li2022}, structurally motivated parameter grouping \cite{Jobman2015, biz18, Chu2019part1, KHALIK2021229901}, or staged estimation
\cite{Jin2018, Fan2020},
although high-dimensional simultaneous fits have also been demonstrated
\cite{FORMAN2012263,Reddy2019,Hallemans2025,Hassanaly2026}.
These studies collectively show that accurate reproduction of terminal voltage does not by itself guarantee unique or physically meaningful parameter recovery \cite{FORMAN2012263,KHALIK2021229901,Escalante2021,Berliner_2021}.

Structural and practical identifiability have also received considerable attention, through structural analysis, Fisher information, confidence-region analysis and Bayesian posterior inference \cite{Schmidt2010,FORMAN2012263,biz18,ait20,Berliner_2021,Laue2021}.
Such analyses remain unevenly integrated into high-dimensional inference problems. Several studies assess sensitivity or fit quality without characterising correlations among the fitted parameters \cite{Jin2018,Reddy2019,Li2022,Xie2026}.
Computational cost remains a further limitation and has motivated reformulated models \cite{Boovaragavan2008}, surrogate forward operators \cite{Hassanaly2024,Hallemans2025}, automatic differentiation \cite{difflib}, batch-parallel optimisation and just-in-time model evaluation \cite{Guo2026}, and amortised neural inference \cite{Hassanaly2026, Zonta2026}.
One recent study brings several of these strands together by combining a differentiable, physics-informed operator surrogate for the SPM with global experimental design, staged differential-evolution fitting and local Fisher information analysis for seven scalar parameters in a synthetic benchmark \cite{Brendel2026}.
Even so, most full-cell inverse studies estimate scalar transport coefficients or, when state dependence is included, infer only an overall scale while keeping the functional form fixed \cite{ait20,Li2022,Hallemans2025,difflib}. Although some specialised functional-inference precedents exist \cite{FORMAN2012263,Ayerbe2022,GUMRUKCUOGLU2026120831}, inference of flexible state-dependent diffusivities together with a broad set of interacting SPMe/DFN parameters remains comparatively uncommon.

The present study is positioned at the intersection of these issues. We first remove exact structural redundancies, but
rather than subsequently selecting a subset of sensitive parameters before inference, as in a number of existing approaches \cite{Jin2018,KHALIK2021229901,Li2022}, we infer a comparatively broad set of interacting SPMe parameter groups, including state-dependent solid and electrolyte diffusivities, from a single dynamic current--voltage dataset.
We then use local sensitivity and Fisher information analysis to assess which parameter combinations the voltage data constrain and which can compensate for one another. This exposes weak directions that prior constraints can mask and shows how the diffusivity parameterisation affects the apparent resolution of the reconstructed functions.

To make physics-based parameter inference from dynamic current--voltage data computationally practical, we use our Artiphy surrogate framework\footnote{\url{http://artiphy.co.uk}} to accelerate evaluation of the SPMe forward map.
We then examine which non-redundant parameter groups and state-dependent diffusivities can be recovered under a prescribed drive-cycle current protocol, which parameters remain weakly constrained or strongly correlated despite an accurate voltage fit, and how parameter recovery is affected in the presence of model discrepancy.

Synthetic DFN data are used as the primary benchmark.
This avoids the same-model ``inverse crime'' \cite{Wirgin2004}: rather than generating and inferring the voltage time series with the same SPMe implementation, we generate data from a more detailed DFN model and infer with an SPMe. The benchmark therefore retains known reference parameters while introducing structured model discrepancy.
This controlled drive-cycle benchmark provides a basis for future tests on experimental current--voltage data, including on-board measurements, where measurement noise and additional model discrepancy must also be addressed.

The paper is organised as follows.
In \S\ref{sec:forward-model}, we introduce the SPMe used for inference. In \S\ref{sec:parameterisation}, we reformulate the model in terms of parameter groups and remove exact structural redundancies.
In \S\ref{sec:synthetic-benchmark}, we describe the synthetic DFN benchmark, including the reference parameter set and current protocol. The inverse formulation is presented in \S\ref{sec:inference}, including the
functional parameterisations, priors and surrogate-accelerated maximum a posteriori (MAP) inference,
and \S\ref{sec:results} reports the resulting parameter estimates.
We then examine the local sensitivity structure and practical identifiability of the inferred parameters in \S\ref{sec:local-identifiability}.
Finally, we summarise our main findings in \S\ref{sec:discussion} and discuss their implications for physics-based parameter inference.
Supporting details on the surrogate, priors and additional sensitivity
diagnostics are provided in the appendices.

\section{Single particle model with electrolyte}
\label{sec:forward-model}
We adopt an isothermal SPMe to compute the terminal voltage from a prescribed applied current and parameter set. It comprises spherical solid diffusion problems in the negative and positive electrodes, a one-dimensional electrolyte concentration problem, and a voltage construction combining open-circuit, reaction, concentration, and Ohmic contributions. To avoid the inconsistent use of anode and cathode in the battery literature, we refer throughout to the negative and positive electrodes, denoted by $\mrn$ and $\mrp$.

\subsection{Solid diffusion in the active material particles}
\label{sec:solid_problem}
For each $k\in\{\mrn,\mrp\}$, the single particle approximation represents the solid concentration by $c_k(r,t)$, independent of the through-cell coordinate $x$, satisfying
\begin{equation}
    \frac{\partial c_k}{\partial t}
    =
    \frac{1}{r^2}\frac{\partial}{\partial r}
    \left(
        r^2 D_k(c_k) \frac{\partial c_k}{\partial r}
    \right),
    \qquad 0<r<R_k,
\label{eq:solid_diff_unscaled}
\end{equation}
subject to boundary and initial conditions:
\begin{equation}
\left.\frac{\partial c_k}{\partial r}\right\vert_{r=0} = 0\,,\qquad
-\left.D_k(c_k)\frac{\partial c_k}{\partial r}\right\vert_{r=R_k} = \frac{j_k(t)}{F}\,,
\qquad
c_k(r,0)=c_{k,0}\,.
\end{equation}
$D_k(c_k)$ is the concentration-dependent solid diffusivity, $R_k$ is the particle radius, $F$ is Faraday's constant, and $j_k(t)$ is the interfacial current density. Uniform distribution of the applied current over the active-particle surface area gives

\begin{equation}
    j_k(t)
    =
    (-1)^{\delta_{k\mrp}}\frac{i_{\mathrm{app}}(t)R_k}{3L_k\varepsilon^{\mathrm{act}}_k}\,,
    \label{eq:flux_to_curr}
\end{equation}
where $i_{\mathrm{app}}$ is the applied areal current density~\footnote{Here $i_{\rm app}$ is an areal current density per projected cell area
(equivalently, per current-collector area), whereas $j_k$ is defined per unit
active-particle surface area.}, $L_k$ is the electrode thickness and $\varepsilon^{\mathrm{act}}_k$ is the active material volume fraction.
The Kronecker delta $\delta_{k\mrp}$ sets the relative sign: positive applied current delithiates the negative electrode ($j_{\rm n}>0$) and lithiates the positive electrode ($j_{\rm p}<0$).

\subsection{Electrolyte concentration dynamics}
\label{sec:liquid_problem}
The electrolyte concentration $c_\mre(x,t)$ is modelled over the full cell thickness $0<x<L$, with $x=0$ and $x=L$ at the negative and positive current collectors, respectively, and $L=L_\mrn+L_\mrs+L_\mrp$, with $L_\mrs$ denoting the separator thickness. Following Ref.~\cite{ric20}, we use the formulation with the anion-flux $\mathcal{F}_-$
\begin{align}
    \varepsilon(x)\frac{\partial c_\mre}{\partial t}
    +
    \frac{\partial \mathcal{F}_-}{\partial x} = 0\,,\qquad
    \mathcal{F}_- = -D_\mre(c_\mre)\mathcal{B}(x)\frac{\partial c_\mre}{\partial x}
    - \frac{1-t^+}{F}i_\mre(x,t)\,,\qquad
    x \in [0,L]\,,
\end{align}
subject to no-flux boundary conditions and a uniform initial concentration
\begin{equation}
\left.\mathcal{F}_-\right\vert_{x=0}=\left.\mathcal{F}_-\right\vert_{x=L}=0\,,\qquad
c_\mre(x,0)=c_{\mre,0}.
\end{equation}
Both $c_\mre$ and $\mathcal F_-$ are continuous at the
electrode--separator interfaces.
This formulation conserves electrolyte inventory, $\int_0^L \varepsilon(x)c_\mre(x,t)\mathrm{d}x=c_{\mre,0}\int_0^L\varepsilon(x)\mathrm{d}x$, and is equivalent, up to notation, to the usual cation-balance formulation (see e.g. Refs.~\cite{mar19, bro22}). We use it because anions are not consumed or produced by the electrode reactions, so their balance contains no intercalation source term. Both $i_\mre$ and $\mathcal F_-$ are homogenised per unit electrode cross-sectional area, rather than local pore-space quantities.

The porosity $\varepsilon(x)$ and the effective transport efficiency factor $\mathcal{B}(x)$  are piecewise constant over the negative electrode, separator, and positive electrode, with
\begin{equation}
\varepsilon(x)= \begin{cases}
                  \varepsilon_\mrn\,,& 0\le x<L_\mrn\,,\\
                  \varepsilon_\mrs\,,& L_\mrn\le x <L-L_\mrp\,,\\
                  \varepsilon_\mrp\,,& L-L_\mrp\le x\le L\,,
                 \end{cases}
\quad\qquad
\mathcal{B}(x) = \begin{cases}
                  \mathcal{B}_\mrn\,,& 0\le x<L_\mrn\,,\\
                  \mathcal{B}_\mrs\,,& L_\mrn\le x <L-L_\mrp\,,\\
                  \mathcal{B}_\mrp\,,& L-L_\mrp\le x\le L\,.
                 \end{cases}
\end{equation}

As a consequence of the uniform interfacial reaction current assumed in Eq.~\eqref{eq:flux_to_curr}, the electrolyte current density $i_\mre(x,t)$ is piecewise linear in the electrodes and constant in the separator,
\begin{equation}
    i_\mre(x,t)=i_{\mathrm{app}}(t)\,
    \begin{cases}
        \dfrac{x}{L_\mrn}, & 0\leq x<L_\mrn, \\
        1 , & L_\mrn\leq x<L-L_\mrp, \\
        \dfrac{L-x}{L_\mrp}, & L-L_\mrp\leq x\leq L\,.
    \end{cases}
\end{equation}
\subsection{Voltage construction}

We use the electrode-averaged voltage construction of Ref.~\cite{mar19}, derived in the combined limit of fast electrolyte transport and high conductivities, with first-order electrolyte concentration corrections. A related construction based on a different asymptotic limit is given in Ref.~\cite{ric20}; both agree closely with the DFN model from which they are derived.

The terminal voltage is decomposed as
\begin{equation}
    V(t)
    =
    U_{\mathrm{eq}}(t)
    - \eta_\mathrm{r}(t)
    - \eta_\mathrm{c}(t)
    - \Delta\phi_\mathrm{e}(t)
    - \Delta\phi_\mathrm{s}(t).
    \label{eq:voltage-decomp}
\end{equation}
where $U_\mathrm{eq}$ is the cell open-circuit potential, $\eta_\mathrm{r}$ and $\eta_\mathrm{c}$ are the reaction and concentration overpotentials, and $\Delta\phi_\mre$ and $\Delta\phi_\mrs$ are the electrolyte and electrode Ohmic losses. The reaction and Ohmic terms have the sign of the applied current, whereas the concentration overpotential depends on the evolving
electrolyte profile.

The open-circuit potential, normally the dominant voltage contribution, depends on the particle surface concentrations:
\begin{equation}
	U_\mathrm{eq} = U_\mrp \left( c_{\mrp} |_{r = R_\mrp} \right) - U_\mrn \left( c_{\mrn} |_{r = R_\mrn} \right).
	\label{eq:Ueq_unscaled}
\end{equation}
The exchange current densities are defined as
\begin{align}
	j_{k,0}(x,t) &= F K_k \left. \sqrt{\frac{c_{\mre}(x,t)}{c_{\mre, 0}} \frac{c_{k}(r,t)}{\cmax} \left(1 - \frac{c_{k}(r,t)}{\cmax} \right)} \right\vert_{r = R_k} \,, \qquad k \in\{\mrn,\mrp\}.
	\label{eq:j0_unscaled}
\end{align}
Here, $K_k$ is a reaction rate constant with SI units of $\mathrm{mol~m^{-2} s^{-1}}$. The SPMe formulation of Ref.~\cite{mar19} uses exchange current densities averaged over the corresponding electrode regions:
\begin{equation}
\bar{j}_{\mrn,0}(t) = \frac{1}{L_\mrn}\int_0^{L_\mrn}j_{\mrn,0}(x,t) dx\,,\qquad
\bar{j}_{\mrp,0}(t) = \frac{1}{L_\mrp}\int_{L-L_\mrp}^{L}j_{\mrp,0}(x,t) dx\,.
\label{eq:j0bar_unscaled}
\end{equation}
The reaction overpotential is then
\begin{equation}
	\eta_\mathrm{r} = \frac{2 R_g T}{F} \left( \arcsinh\left( \frac{j_\mrn(t)}{\bar{j}_{\mrn,0}(t)} \right)  - \arcsinh\left( \frac{j_\mrp(t)}{\bar{j}_{\mrp,0}(t)} \right) \right)\,,
	\label{eq:etar_unscaled}
\end{equation}
where $R_g$ is the universal gas constant and $T$ is the prescribed temperature.
This is a symmetric Butler--Volmer relation with transfer coefficients $1/2$, using the exchange-current normalisation of Ref.~\cite{mar19}.

The concentration overpotential obtained from the electrolyte solution is
\begin{equation}
	\eta_\mathrm{c} = \frac{2 R_g T\,(1-t^+) }{F\,c_{\mre, 0}} \left( \frac{1}{L _\mrn} \int_{0}^{L_\mrn} c_\mre (x,t) \dd x  -  \frac{1}{L _\mrp} \int_{L - L_\mrp}^{L} c_\mre (x,t) \dd x \right)\,.
	\label{eq:etac_unscaled}
\end{equation}

In Ref.~\cite{mar19}, the electrolyte Ohmic loss depends only on scalar parameters and the areal current density:
\begin{equation}
	\Delta \phi_\mre = \frac{i_\mathrm{app}}{\sigma_\mre(c_{\mre, 0})}\left(\frac{L_\mrn}{3\,\mathcal{B}_{\mrn}}+\frac{L_\mrs}{\mathcal{B}_{\mrs}}+\frac{L_\mrp}{3\,\mathcal{B}_{\mrp}}\right)\,,
	\label{eq:DFe_unscaled}
\end{equation}
where the concentration-dependent DFN conductivity $\sigma_\mre(c_\mre)$ is projected to its value $\sigma_\mre(c_{\mre,0})$ at the reference concentration.
Finally, the solid-phase Ohmic loss, often small in well-designed electrodes, is retained for completeness:
\begin{equation}
	\Delta \phi_\mrs = \frac{i_\mathrm{app}}{3} \left( \frac{L _\mrp}{\sigma_\mrp} + \frac{L _\mrn}{\sigma_\mrn}\right).
	\label{eq:DFs_unscaled}
\end{equation}

\section{Parameter grouping and exact structural redundancies}
\label{sec:parameterisation}

In this section, we rescale the SPMe to remove exact algebraic redundancies: distinct dimensional parameter sets give the same rescaled model when they share the combinations appearing in its equations. The resulting groups provide a non-redundant parametrisation under the modelling and known-input assumptions stated below. Removing these redundancies is necessary, but not sufficient, for structural identifiability from terminal-voltage data. \S\ref{sec:local-identifiability} addresses a separate question: whether the voltage response to the chosen drive cycle locally distinguishes the remaining parameters.
Our approach builds on the lumped-parameter reformulation of the DFN model by Jobman et al. \cite{Jobman2015}, the structural-identifiability analysis of the SPM by Bizeray et al. \cite{biz18}, and the recent grouped-parameter SPMe of Hallemans et al. \cite{Hallemans2025}, adapting these ideas to an SPMe with concentration-dependent transport functions.

\subsection{Scalings and normalisations}
\label{sec:scalings}

We first scale the spatial coordinates, concentrations and anion flux:
\begin{subequations}
\begin{align}
c_k &= \cmax x_k\,, \qquad r_k = R_k \scriptr_k\,, \qquad k\in\{\mrn,\mrp\}\,,
\label{eq:dimless_solid}\\
c_\mre &= c_{\mre,0}\hat c_\mre\,, \qquad x=L\hat x\,, \qquad
\mathcal F_- = \frac{c_{\mre,0}\mathcal B_\mrs D_{\mre,\rm ref}}{L}\mathcal F_-^\star\,.
\label{eq:liquid_scalings}
\end{align}
\end{subequations}
Time, current densities and voltage remain dimensionful.

The diffusivity functions are separated into scalar scales and dimensionless concentration dependence:
\begin{equation}
D_k(c_k)=D_{k,\rm mean}D_k^\star(x_k)\,, \qquad
D_\mre(c_\mre)=D_{\mre,\rm ref}D_\mre^\star(\hat c_\mre)\,.
\label{eq:diffusivity_scalings}
\end{equation}
The scales and normalisations are defined by
\begin{subequations}
\begin{align}
\int_0^{\cmax}\log\left(\frac{D_k(c)}{D_{k,\rm mean}}\right)\dd c &= 0\,,
\label{eq:mean_diff}\\
D_{\mre,\rm ref} &= D_\mre(c_{\mre,0})\,,
\label{eq:Deref_def}\\
D_\mre^\star(1) &= 1\,.
\label{eq:De_normalisation}
\end{align}
\end{subequations}
Thus $D_{k,\rm mean}$ sets the overall solid diffusivity scale and $D_k^\star(x_k)$ its concentration dependence. The fixed stoichiometry interval $[0,1]$ makes a unit geometric mean convenient for the Fourier representation used in \S\ref{sec:inference}. For the electrolyte, normalisation at the prescribed initial concentration provides a reference without choosing a concentration interval over which to average.

The electrode OCP functions are also expressed in terms of stoichiometry:
\begin{equation}
U_k^{(x)}(x_k)=U_k(\cmax x_k)\,.
\label{eq:ocp_stoichiometry}
\end{equation}
For notational brevity, in the remainder of the paper we use $U_k(x_k)$ as shorthand for $U_k^{(x)}(x_k)$; the potential itself remains dimensionful.

The regional geometry and material properties are expressed as
\begin{equation}
\lambda_i=\frac{L_i}{L}\,,\qquad
\hat\varepsilon_i=\frac{\varepsilon_i}{\varepsilon_\mrs}\,,\qquad
\hat{\mathcal B}_i=\frac{\mathcal B_i}{\mathcal B_\mrs}\,,\qquad
i\in\{\mrn,\mrs,\mrp\}\,,
\label{eq:regional_scalings}
\end{equation}
with $\lambda_\mrs=1-\lambda_\mrn-\lambda_\mrp$ and $\hat\varepsilon_\mrs=\hat{\mathcal B}_\mrs=1$. Hence $\varepsilon(x)=\varepsilon_\mrs\hat\varepsilon(\hat x)$ and $\mathcal B(x)=\mathcal B_\mrs\hat{\mathcal B}(\hat x)$, where
\begin{subequations}
\begin{align}
\begin{split}
\hat\varepsilon(\hat{x})&= \begin{cases}
                  \hat\varepsilon_\mrn\,,& 0\le \hat{x}<\lambda_\mrn\,,\\
                  1\,,& \lambda_\mrn\le \hat{x} <1-\lambda_\mrp\,,\\
                  \hat\varepsilon_\mrp\,,& 1-\lambda_\mrp\le \hat{x}\le 1\,,
                 \end{cases}
\end{split}\\
\begin{split}
\hat{\mathcal{B}}(\hat{x}) &= \begin{cases}
                  \hat{\mathcal{B}}_\mrn\,,& 0\le \hat{x}<\lambda_\mrn\,,\\
                  1\,,& \lambda_\mrn\le \hat{x} <1-\lambda_\mrp\,,\\
                  \hat{\mathcal{B}}_\mrp\,,& 1-\lambda_\mrp\le \hat{x}\le 1\,,
                 \end{cases}
\end{split}
\end{align}
\label{eq:regional_profiles}
\end{subequations}

For this study, we assume the Bruggeman relation with a fixed exponent of $3/2$, the value obtained from an idealised model of ion transport around spherical particles \cite{bruggeman35}:
\begin{equation}
\mathcal B_i=\varepsilon_i^{3/2}\,,\qquad
\hat{\mathcal B}_i=\hat\varepsilon_i^{3/2}\,,\qquad i\in\{\mrn,\mrs,\mrp\}\,.
\label{eq:bruggeman_rescaled}
\end{equation}

\subsection{Rescaled SPMe}
\label{sec:scaled_dyn}
\label{sec:scaled_v_assembly}

We now state the rescaled dynamical problem and voltage construction. The scalar groups appearing below are defined together in \S\ref{sec:reduced_groups}.

Using the scalings above and the relation between interfacial flux and applied current density \eqref{eq:flux_to_curr}, the solid diffusion equations become
\begin{align}
\tau_k\pdv{x_k}{t} &= \frac{1}{\scriptr^2}\pdv{}{\scriptr}\left(\scriptr^2D_k^\star(x_k)\pdv{x_k}{\scriptr}\right)\,,\nonumber\\
\left.\pdv{x_k}{\scriptr}\right\vert_{\scriptr=0} &= 0\,,\nonumber\\
\left.-D_k^\star(x_k)\pdv{x_k}{\scriptr}\right\vert_{\scriptr=1} &= (-1)^{\delta_{k\mrp}}\frac{\tau_k}{3\mathcal Q_k}i_{\rm app}(t)\,,\nonumber\\
x_k\big\vert_{t=0} &= x_{k,0}\,,\qquad k\in\{\mrn,\mrp\}\,.
\label{eq:solid_dynamics_rescaled}
\end{align}
Similarly, the rescaled electrolyte concentration dynamics are
\begin{align}
\tau_\mre\hat\varepsilon(\hat x)\frac{\partial\hat c_\mre}{\partial t}
+\frac{\partial\mathcal F_-^\star}{\partial\hat x} &= 0\,,\nonumber\\
\mathcal F_-^\star &= -D_\mre^\star(\hat c_\mre)\hat{\mathcal B}(\hat x)\frac{\partial\hat c_\mre}{\partial\hat x}
-\frac{(1-t^+)\tau_\mre}{\mathcal Q_\mre}i_\mre(\hat x,t)\,,\nonumber\\
\mathcal F_-^\star\big\vert_{\hat x=0} &= \mathcal F_-^\star\big\vert_{\hat x=1}=0\,,\nonumber\\
\hat c_\mre\big\vert_{t=0} &= 1\,,
\label{eq:liquid_dynamics_rescaled}
\end{align}
with continuity of $\hat c_\mre$ and $\mathcal F_-^\star$ at the electrode--separator interfaces. The electrolyte current density is
\begin{equation}
i_\mre(\hat x,t)=i_{\rm app}(t)
\begin{cases}
\dfrac{\hat x}{\lambda_\mrn}, & 0\le\hat x<\lambda_\mrn\,,\\
1, & \lambda_\mrn\le\hat x<1-\lambda_\mrp\,,\\
\dfrac{1-\hat x}{\lambda_\mrp}, & 1-\lambda_\mrp\le\hat x\le1\,.
\end{cases}
\label{eq:electrolyte_current_rescaled}
\end{equation}

The terminal voltage retains the decomposition
\begin{equation}
V(t)=U_{\rm eq}(t)-\eta_{\rm r}(t)-\eta_{\rm c}(t)-\Delta\phi_\mre(t)-\Delta\phi_\mrs(t)\,.
\label{eq:voltage_rescaled}
\end{equation}
The open-circuit potential is now written in terms of the surface stoichiometries:
\begin{equation}
U_{\rm eq}=U_\mrp\left(x_\mrp\big\vert_{\scriptr=1}\right)-U_\mrn\left(x_\mrn\big\vert_{\scriptr=1}\right)\,.
\label{eq:Ueq_scaled}
\end{equation}
For the reaction overpotential, we use Eqs.~\eqref{eq:j0_unscaled}--\eqref{eq:j0bar_unscaled} to define exchange current densities per geometric area,
\begin{equation}
i_{k,0}=3\frac{L_k\varepsilon_k^{\rm act}}{R_k}\bar j_{k,0}\,,
\end{equation}
which become
\begin{equation}
\begin{aligned}
i_{\mrn,0} &= \left.\frac{i^{\rm ref}_{\mrn,0}\sqrt{x_\mrn(1-x_\mrn)}}{\lambda_\mrn}
\int_0^{\lambda_\mrn}\sqrt{\hat c_\mre}\dd\hat x\right\vert_{\scriptr=1}\,,\\
i_{\mrp,0} &= \left.\frac{i^{\rm ref}_{\mrp,0}\sqrt{x_\mrp(1-x_\mrp)}}{\lambda_\mrp}
\int_{1-\lambda_\mrp}^{1}\sqrt{\hat c_\mre}\dd\hat x\right\vert_{\scriptr=1}\,.
\end{aligned}
\label{eq:j0bar_scaled}
\end{equation}
The reaction and concentration overpotentials are then
\begin{align}
\eta_{\rm r} &= \nu\left[\arcsinh\left(\frac{i_{\rm app}(t)}{i_{\mrn,0}(t)}\right)
+\arcsinh\left(\frac{i_{\rm app}(t)}{i_{\mrp,0}(t)}\right)\right]\,,
\label{eq:etar_scaled}\\
\eta_{\rm c} &= \nu(1-t^+)\left(\frac{1}{\lambda_\mrn}\int_0^{\lambda_\mrn}\hat c_\mre\dd\hat x
-\frac{1}{\lambda_\mrp}\int_{1-\lambda_\mrp}^{1}\hat c_\mre\dd\hat x\right)\,.
\label{eq:etac_scaled}
\end{align}
Finally, the Ohmic drops are proportional to the applied current density:
\begin{align}
\Delta\phi_\mre &= \mathcal R_\mre i_{\rm app}\,,
\label{eq:DFe_scaled}\\
\Delta\phi_\mrs &= \mathcal R_\mrs i_{\rm app}\,.
\label{eq:DFs_scaled}
\end{align}

\subsection{Reduced parameter groups}
\label{sec:reduced_groups}

The scalar groups introduced in the rescaled equations are
\begin{subequations}
\label{eq:reduced_parameter_definitions}
\begin{align}
\tau_k &= \frac{R_k^2}{D_{k,\rm mean}}\,,\qquad
\mathcal Q_k=F\varepsilon_k^{\rm act}L_k\cmax\,,
\label{eq:tauk_def}\\
\tau_\mre &= \frac{\varepsilon_\mrs L^2}{\mathcal B_\mrs D_{\mre,\rm ref}}\,,
\label{eq:taue_def}\\
\mathcal Q_\mre &= F c_{\mre,0}L\varepsilon_\mrs\,,
\label{eq:Qe_def}\\
i^{\rm ref}_{k,0} &= \frac{3FK_kL_k\varepsilon_k^{\rm act}}{R_k}\,,\qquad
\nu=\frac{2R_gT}{F}\,,
\label{eq:kinetic_scales}\\
\mathcal R_\mre &= \frac{1}{\sigma_\mre(c_{\mre,0})}
\left(\frac{L_\mrn}{3\mathcal B_\mrn}+\frac{L_\mrs}{\mathcal B_\mrs}+\frac{L_\mrp}{3\mathcal B_\mrp}\right)\,,
\label{eq:fancyR_e}\\
\mathcal R_\mrs &= \frac{1}{3}\left(\frac{L_\mrn}{\sigma_\mrn}+\frac{L_\mrp}{\sigma_\mrp}\right)\,,\qquad
\mathcal R_{\rm ohm}=\mathcal R_\mre+\mathcal R_\mrs\,.
\label{eq:fancyR_s}
\end{align}
\end{subequations}
Here $\tau_k$ is the mean solid diffusion time and $\mathcal Q_k$ the maximum areal lithium inventory. The electrolyte scales $\tau_\mre$ and $\mathcal Q_\mre$ are referenced to separator properties: $\mathcal Q_\mre$ is the charge-equivalent salt inventory per unit area for a domain of thickness $L$ and uniform porosity $\varepsilon_\mrs$. The choice of separator values as reference for the electrolyte variables is discussed in \ref{app:electrolyte_coordinates}.

The quantities $i_{k,0}$ and $i^{\rm ref}_{k,0}$ are, respectively, the exchange current density and its reference scale, both per projected current-collector area. The quantities $\mathcal R_\mre$ and $\mathcal R_\mrs$ are area-specific Ohmic resistances, and $\nu$ is the thermal voltage scale.

\subsection{Exact redundancies and assumptions}
\label{sec:parameter_groups_rescaled}

For each electrode, Eq.~\eqref{eq:solid_dynamics_rescaled} depends on two scalar groups and one functional parameter:
\begin{equation}
\tau_k\,,\qquad\mathcal Q_k\,,\qquad D_k^\star(x_k)\,,\qquad k\in\{\mrn,\mrp\}\,.
\end{equation}
The particle radius and diffusivity scale enter through $\tau_k$ and cannot be separated using the particle diffusion equation alone. Similarly, the factors in $\mathcal Q_k$ enter that equation only through their product. The group $\mathcal Q_k$ sets the current-to-stoichiometry scale, while $D_k^\star(x_k)$ describes the concentration dependence of the diffusivity.

Under the Bruggeman relation \eqref{eq:bruggeman_rescaled}, Eq.~\eqref{eq:liquid_dynamics_rescaled} depends on six scalar groups and one functional parameter:
\begin{equation}
\lambda_\mrn\,,\qquad\lambda_\mrp\,,\qquad
\hat\varepsilon_\mrn\,,\qquad\hat\varepsilon_\mrp\,,\qquad
\tau_\mre\,,\qquad\frac{\mathcal Q_\mre}{1-t^+}\,,\qquad
D_\mre^\star(\hat c_\mre)\,.
\label{eq:electrolyte_lumped}
\end{equation}
These six scalar coordinates are in one-to-one correspondence with the six independent regional groups; the construction is given in \ref{app:electrolyte_coordinates}. A closely related electrolyte parametrisation is used by Hallemans et al. \cite{Hallemans2025}, in terms of relative electrode thicknesses and porosities, a reference electrolyte capacity, and regional electrolyte diffusion time scales.

The voltage construction adds the kinetic scales $i^{\rm ref}_{k,0}$, the thermal voltage scale $\nu$, and the Ohmic resistance groups. The two Ohmic losses enter terminal voltage only through their sum,
\begin{equation}
\Delta\phi_\mre+\Delta\phi_\mrs=\mathcal R_{\rm ohm}i_{\rm app}\,.
\end{equation}
Consequently, $\mathcal R_\mre$ and $\mathcal R_\mrs$ cannot be recovered separately from terminal-voltage data under this formulation. The concentration overpotential \eqref{eq:etac_scaled} contains an additional factor $1-t^+$, removing the particular algebraic redundancy between $\mathcal Q_\mre$ and $t^+$ in the electrolyte dynamics. We therefore retain $\mathcal Q_\mre$ and $t^+$ separately, without claiming that this establishes their input--output identifiability.

Full-cell equilibrium voltage observes only the difference $U_\mrp(x_\mrp)-U_\mrn(x_\mrn)$, so it cannot by itself determine the two electrode OCP functions. In this work, we treat $U_\mrn$ and $U_\mrp$ as known from prior electrode characterisation. Even with known OCP functions, a single equilibrium-voltage value does not determine both initial stoichiometries. We prescribe $x_{\mrn,0}$ and $x_{\mrp,0}$ from the synthetic DFN benchmark. For a physical cell, prescribing these values would require additional initial-state or alignment information.

Prescribing the initial stoichiometries does not fix the electrode-capacity ratio. An independently established full-cell alignment, such as $x_\mrp=x_\mrp(x_\mrn)$, would constrain $\mathcal Q_\mrp/\mathcal Q_\mrn$ \cite{KHALIK2021229901}. We impose no such constraint, so $\mathcal Q_\mrn$ and $\mathcal Q_\mrp$ remain independent parameter groups.

Unlike Bizeray et al.~\cite{biz18}, who analyse the input--output structural identifiability of an SPM linearised around equilibrium through its transfer function, we do not attempt an analogous small-signal analysis. Such an analysis would characterise only the parameter combinations visible in the local small-signal dynamics and the local values of the concentration-dependent transport functions, rather than the nonlinear response under the large-amplitude current protocol considered here. We assess the local distinguishability of the remaining finite-dimensional parameters for the chosen protocol in \S\ref{sec:local-identifiability}.

Subject to the Bruggeman relation, prescribed temperature, known electrode OCP functions and prescribed initial state, the parameter groups remaining after removal of these exact redundancies are summarised in Table~\ref{tab:structurally_id_params}.

\begin{table}[h!]
\newcolumntype{Y}{
    >{\raggedright\arraybackslash
      \hangindent=1em
      \hangafter=1}X
}
\centering
\small
\setlength{\tabcolsep}{4pt}

\begin{tabularx}{\textwidth}{
    @{}
    l
    l
    Y
    >{\raggedright\arraybackslash}X
    c
    @{}
}
\toprule
Parameter & SI Units & Physical meaning & Model contribution &  Count \\
\midrule
$\tau_k$ & \unit{\second} & Solid diffusion time scale & Solid dynamics $\to U_{\rm eq},\eta_{\rm r}$ &  2 S\\
$\mathcal{Q}_k$ & \unit{\coulomb\per\metre\squared} &Maximum areal lithium inventory &  Solid dynamics $\to U_{\rm eq},\eta_{\rm r}$& 2 S\\
$i^{\rm ref}_{k,0}$ & \unit{\ampere\per\metre\squared} & Geometric exchange current density scale & $\eta_{\rm r}$ & 2 S \\
$\tau_\mre$ &\unit{\second}& Separator-referenced diffusion time in the electrolyte & Electrolyte dynamics $\to \eta_{\rm c},\eta_{\rm r}$ & 1 S\\
$\mathcal{Q}_\mre$ &\unit{\coulomb\per\metre\squared}& Separator-referenced areal electrolyte salt-inventory scale&  Electrolyte dynamics $\to  \eta_{\rm c}$, $\eta_{\rm r}$ & 1 S\\
$t^+$ & \textendash & Transference number & Electrolyte dynamics and $\eta_{\rm c}$& 1 S\\
$\mathcal{R}_{\rm ohm}$& \unit{\ohm\,\metre\squared} & Combined area-specific Ohmic resistance& $\Delta\phi_\mre,  \Delta\phi_\mrs$ & 1 S\\
$\lambda_k$ &\textendash & Electrode thickness fraction &  Electrolyte dynamics $\to \eta_{\rm c},\eta_{\rm r}$& 2 S\\
$\hat\varepsilon_k$ &\textendash & Porosity relative to separator & Electrolyte dynamics $\to \eta_{\rm c},\eta_{\rm r}$& 2 S\\
\midrule
$D^\star_k(x_k)$ & \textendash& Solid-state diffusivity with unit geometric mean & Solid dynamics $\to U_{\rm eq},\eta_{\rm r}$ & 2 F\\
$D^\star_\mre(\hat{c}_\mre)$ & \textendash& Electrolyte diffusivity normalised by its value at the initial concentration & Electrolyte dynamics $\to \eta_{\rm c},\eta_{\rm r}$ & 1 F\\
\midrule
{\bf TOTAL} &&&&14 S, 3 F\\
\bottomrule
\end{tabularx}

\caption{
Scalar and functional parameter groups in the non-redundant SPMe parameterisation used in this study, under the Bruggeman relation, prescribed temperature, known electrode OCP functions and prescribed initial state.
Here S denotes a scalar parameter group and F denotes an unknown function. Quantities indexed by $k$ are counted over $k\in\{\mrn,\mrp\}$. The penultimate column indicates the model component through which each parameter group enters the voltage expression; it should not be read as assuming that the voltage components are separately measured.
}
\label{tab:structurally_id_params}
\end{table}

\section{Synthetic DFN benchmark}
\label{sec:synthetic-benchmark}
To assess parameter recovery under model discrepancy, we use a synthetic benchmark with known data-generating DFN parameters. The benchmark voltage is generated using Dandeliion \cite{dande}.
This provides a controlled inverse problem while avoiding the overly favourable case in which the inference model is identical to the data-generating model. The benchmark voltage is intentionally left noise-free so that the effects of the structured SPMe--DFN discrepancy on the voltage residuals and inferred parameter values can be examined without dependence on a particular noise realisation.

The parameter set used for the simulation is listed in Table~\ref{tab:benchmark-parameters}. We use a synthetic but physically plausible parameter set, with a layered-oxide-like positive electrode and a graphite-like negative electrode. The negative-electrode open-circuit potential uses the Dandeliion fit \cite{dandeweb} to graphite data from Ref.~\cite{ecker15}. The positive-electrode open-circuit potential uses the analytical Dandeliion fit whose coefficients match those reported in the supplementary material of Ref.~\cite{zulke21}. The electrolyte diffusivity and conductivity are the fits reported in Ref.~\cite{ecker15}. The solid-state diffusivities are synthetic positive functions chosen to lie within ranges reported for lithium-ion electrode materials \cite{Wang_2022}, while providing nontrivial concentration dependence for the inverse problem.
Inactive solid phases, such as binder and conductive additives, are not represented explicitly, so $\varepsilon^{\rm act}_k=1-\varepsilon_k$.
%
\begin{sidewaystable}[p]
\centering
\begin{threeparttable}
\begin{tabular}{llcl}
\toprule
Quantity & Symbol & Value & Unit\\
\midrule
Electrode cross-sectional area & \textendash & 0.3168 & \unit{\metre\squared}\\
Negative electrode thickness & $L_\mrn$ & 70 & \unit{\micro\metre}  \\
Separator thickness & $L_\mrs$ & 20 & \unit{\micro\metre} \\
Positive electrode thickness & $L_\mrp$ & 60 & \unit{\micro\metre} \\
Negative particle radius & $R_\mrn$ & 5 & \unit{\micro\metre}\\
Positive particle radius & $R_\mrp$ & 4 & \unit{\micro\metre}\\
Negative electrode porosity & $\varepsilon_\mrn$ &0.3 & \textendash\\
Positive electrode porosity & $\varepsilon_\mrp$ &0.32 & \textendash\\
Separator porosity & $\varepsilon_\mrs$ &0.4 & \textendash\\
Electrolyte reference concentration & $c_{\mre,0}$ & 1000 & \unit{\mol\,\metre^{-3}}\\
Negative maximum concentration & $c_{\mrn,\max}$ & 48000 &\unit{\mol\,\metre^{-3}} \\
Positive maximum concentration & $c_{\mrp,\max}$ & 51000 & \unit{\mol\,\metre^{-3}}\\
Transference number & $t^+$ & 0.38 & \textendash\\
Temperature & $T$ & 298.15 & \unit{\kelvin}\\
Transport efficiency in the negative electrode & $\mathcal{B}_\mrn$ & 0.164317 & \textendash\\
Transport efficiency in the positive electrode & $\mathcal{B}_\mrp$ & 0.181019 & \textendash\\
Transport efficiency in the separator & $\mathcal{B}_\mrs$ & 0.252982& \textendash\\
Reaction rate constant in the negative electrode \tnotex{tn:c}
& $K_\mrn$ & $7.58947\times 10^{-5}$ & \unit{\mol\,\metre^{-2}\,\second^{-1}}\\
Reaction rate constant in the positive electrode
& $K_\mrp$ & $9.67657\times 10^{-5}$ & \unit{\mol\,\metre^{-2}\,\second^{-1}}\\
Negative electrode conductivity& $\sigma_\mrn$ & 100 & \unit{\siemens\,\metre^{-1}}\\
Positive electrode conductivity& $\sigma_\mrp$ & 8 & \unit{\siemens\,\metre^{-1}}\\
Conductivity of the electrolyte & $\sigma_\mre(c_\mre)$ & $0.1726 + 1.7919\times10^{-3}\,c_\mre - 1.2983\times10^{-6}\,c_\mre^2 + 0.2667\times10^{-9}\,c_\mre^3 $ & \unit{\siemens\,\metre^{-1}}\\
\addlinespace[0.2em]
Diffusivity in negative electrode particles& $D_\mrn(x_\mrn)$ &
$2\times 10^{-15} \,\exp\left[ -6 \,(x_\mrn-0.5)^2\right] \, \left(2 + \sin(19 \,x_\mrn^2)\right)$
& \unit{\metre\squared\,\second^{-1}} \\
\addlinespace[0.2em]
Diffusivity in positive electrode particles& $D_\mrp(x_\mrp)$ &
\begin{tabular}[t]{@{}l@{}}
$3\times 10^{-15}
 \exp\!\left[-9\,(x_\mrp-0.5)^2\right]$\\
${}\times\Bigl[1
 +0.58\,\tanh\!\left[9\,(x_\mrp-0.35)\right]
 +0.20\,\tanh\!\left[14\,(x_\mrp-0.75)\right]
 \Bigr]$
\end{tabular}
& \unit{\metre\squared\,\second^{-1}} \\
\addlinespace[0.2em]
Diffusivity of the electrolyte & $D_\mre(c_\mre)$ &
\begin{tabular}[t]{@{}l@{}}
$2.646\times10^{-7}\,c_\mre^{-1}$\\
$\times\left(0.1726 + 1.7919\times10^{-3}\,c_\mre - 1.2983\times10^{-6}\,c_\mre^2 + 0.2667\times10^{-9}\,c_\mre^3\right) $
\end{tabular}
& \unit{\metre\squared\,\second^{-1}}
\\
\addlinespace[0.2em]
Negative electrode equilibrium potential & $U_\mrn(x_\mrn)$ &
\begin{tabular}[t]{@{}l@{}}
$0.14003 + 0.7165 \, \exp(-369.03 \, x_\mrn) + 0.1219 \, \exp[-35.648 \, (x_\mrn - 0.053095)]$\\
$- 0.018919\, \tanh[21.197 (x_\mrn - 0.19618)] - 0.016964  \tanh[27.137 (x_\mrn - 0.31283)]$\\
$- 0.019931 \tanh[28.57 (x_\mrn - 0.61422)] - 0.93115 \exp[36.328 (x_\mrn - 1.1074)]$
\end{tabular}
& \unit{\volt}
\\
\addlinespace[0.2em]
Positive electrode equilibrium potential & $U_\mrp(x_\mrp)$ &
\begin{tabular}[t]{@{}l@{}}
$4.17032 - 1.30202\,x_\mrp - 0.214712 \tanh(23.01 (x_\mrp + 0.00350287))$\\
$+ 2.45808 \tanh[2.90232 (x_\mrp  - 0.215657)] - 1.26644 \tanh[4.30574 (x_\mrp  - 0.329193)]$\\
$- 0.40112 \tanh[9.07273 (x_\mrp  - 0.148644)] - 0.0532656 \exp[47.0417  (x_\mrp  - 0.95703)]$
\end{tabular}
& \unit{\volt}
\\
\bottomrule
\end{tabular}
\begin{tablenotes}
\footnotesize
\item[a] \label{tn:c} In Dandeliion, a different convention for the reaction rate
constant is adopted, where $k_k$ has units of
\unit{\metre^{5/2}\,\mol^{-1/2}\,\second^{-1}} \cite{dande}. It is related
to our convention by $K_k = 2\,k_k \sqrt{c_{\mre,0}}\,\cmax$.
\end{tablenotes}
\caption{Scalar and functional parameters used to generate the DFN benchmark data. The full DFN simulation can be accessed through the development version of Dandeliion here: \href{https://development.dandeliion.com/simulation/5632f5fd-3ecc-4c02-912a-9663eff75a7b}{\texttt{https://development.dandeliion.com/simulation/5632f5fd-3ecc-4c02-912a-9663eff75a7b}}.}
\label{tab:benchmark-parameters}
\end{threeparttable}
\end{sidewaystable}
%

\begin{figure}[p]
\centering
\includegraphics[width=\linewidth]{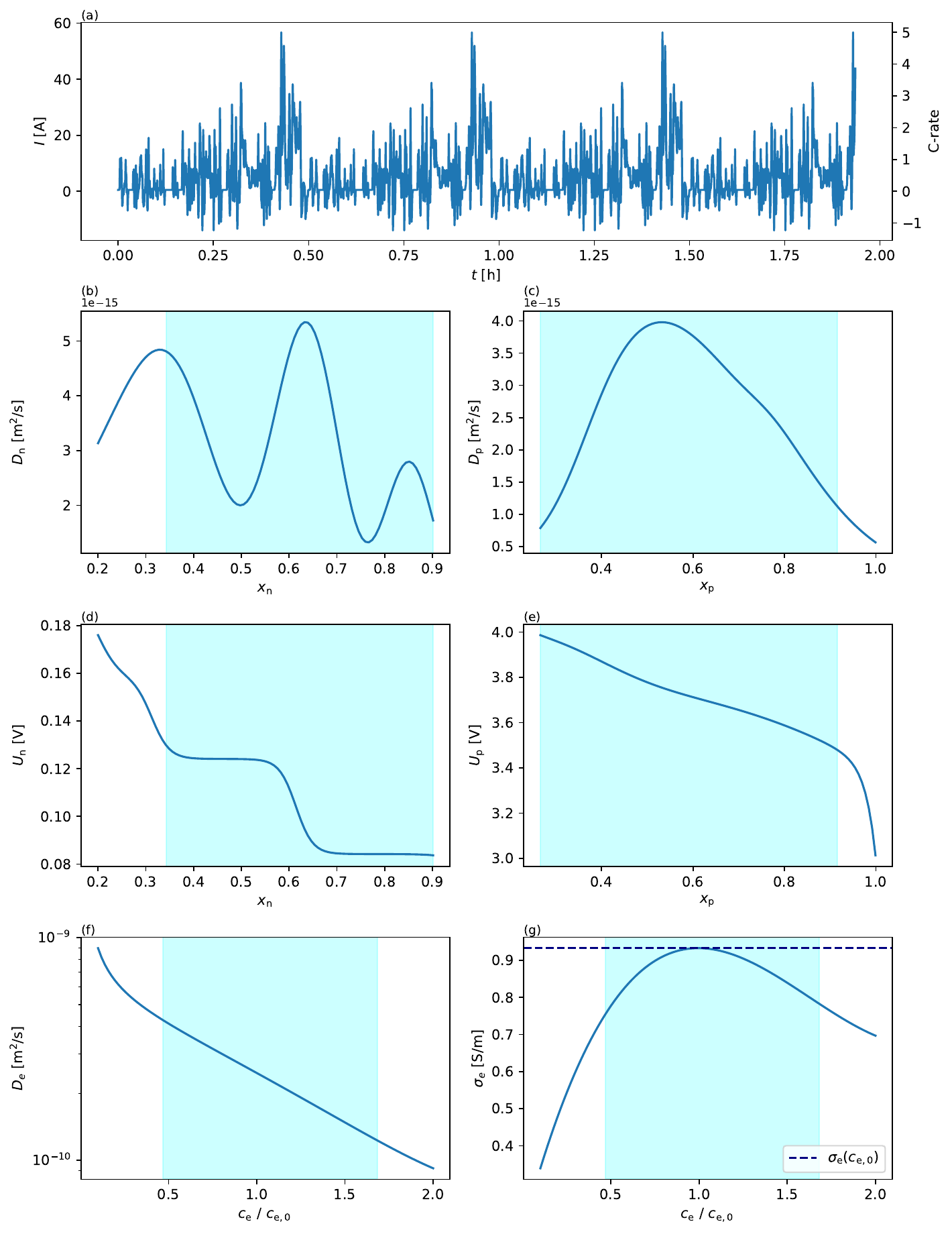}
\caption{Summary of the current protocol and functional parameters used in the synthetic DFN benchmark. (a) Two-hour WLTP-like current profile, shown using the discharge-positive convention adopted in this work, as total cell current on the left axis and the corresponding C-rate on the right axis. (b, c) Solid diffusivities in the negative and positive electrodes, respectively. (d, e) Equilibrium potentials in the negative and positive electrodes. (f, g) Electrolyte diffusivity and conductivity as functions of the normalised electrolyte concentration. The shaded regions indicate the stoichiometry or electrolyte-concentration ranges visited during the benchmark simulation. The dashed line in (g) denotes the constant value $\sigma_\mre(c_{\mre,0})$ in the corresponding SPMe model.}
\label{fig:benchmark-summary}
\end{figure}

The benchmark functional parameters are plotted in Fig.~\ref{fig:benchmark-summary}, together with the ranges of stoichiometry and electrolyte concentration visited during the DFN simulation. In panel (g), the reference conductivity $\sigma_\mre(c_{\mre,0})$, defined after Eq.~\eqref{eq:DFe_unscaled}, lies above the concentration-dependent conductivity over most of the range visited. Since the reduced resistance $\mathcal R_\mre$, defined in Eq.~\eqref{eq:fancyR_e}, is evaluated using this reference value, the corresponding SPMe is expected to underestimate the electrolyte Ohmic loss $\Delta\phi_\mre$.
For comparison with the inferred SPMe, the dimensionful DFN benchmark parameters are projected onto the reduced SPMe coordinates introduced in \S\ref{sec:parameterisation}. The resulting projected DFN benchmark values are listed in Table~\ref{tab:benchmark_structurally_id}. We stress that these are derived reference coordinates, not parameters of an SPMe used to generate the data.
\begin{table}[htb]
\centering
\begin{tabular}{ll}
\toprule
Parameter & Value \\
\midrule
$\tau_\mrn$ & $2.86$ \unit{\hour} \\
$\mathcal{Q}_\mrn$ & $63.04$ \unit{\ampere \hour\per\metre\squared} \\
$i^{\rm ref}_{\mrn,0}$ & $215.29$ \unit{\ampere\per\metre\squared} \\
\midrule
$\tau_\mrp$ & $3.69$ \unit{\hour} \\
$\mathcal{Q}_\mrp$ & $55.77$ \unit{\ampere \hour\per\metre\squared} \\
$i^{\rm ref}_{\mrp,0}$ & $285.70$ \unit{\ampere\per\metre\squared} \\
\midrule
$\tau_\mre$ &$144.12 \unit{\second}$\\
$\mathcal{Q}_\mre$ &$1.61$ \unit{\ampere \hour\per\metre\squared}\\
$t^+$ & $0.38$\\
\midrule
$\mathcal{R}_{\rm ohm}$ & $3.58\times10^{-4}$ \unit{\ohm\,\metre\squared} \\
\midrule
$\lambda_\mrn$ &$0.47$ \\
$\lambda_\mrp$ &$0.40$ \\
\midrule
$\hat\varepsilon_\mrn$ &$0.75$ \\
$\hat\varepsilon_\mrp$ &$0.80$ \\
$\hat{\mathcal{B}}_\mrn$ &$0.65$ {\rm ~~~(derived)} \\
$\hat{\mathcal{B}}_\mrp$ &$0.72$ {\rm ~~~(derived)} \\
\midrule
$\nu$ & $0.051$ \unit{\volt}   {\rm ~~(fixed)} \\
\bottomrule
\end{tabular}
\caption{
Projected DFN benchmark values in the grouped SPMe scalar coordinates, together with derived quantities.
The transport-efficiency ratios $\hat{\mathcal{B}}_i$ and thermal voltage scale $\nu$ are derived from the prescribed porosities and temperature, respectively.
The underlying functional parameters are given in
Table~\ref{tab:benchmark-parameters} and normalised as described in
\S\ref{sec:parameterisation}.}
\label{tab:benchmark_structurally_id}
\end{table}

The benchmark current is a Worldwide Harmonised Light vehicles Test Procedure (WLTP) dynamic profile sampled at $1\,\unit{\hertz}$. After conversion from the Dandeliion sign convention, we report the discharge-positive total current as $I(t)$ and define the corresponding areal current density as $i_{\rm app}(t)=I(t)/A$, where $A$ is the electrode cross-sectional area listed in Table~\ref{tab:benchmark-parameters}.
The cell is initialised at the prescribed electrode stoichiometries and evolved isothermally at room temperature. These and the remaining protocol details are summarised in Table~\ref{tab:benchmark-current-summary}.
The resulting current trace and C-rate are shown
in panel (a) of Fig.~\ref{fig:benchmark-summary}.
We use a dynamic protocol rather than a constant-current discharge because
the voltage sensitivity to transport and kinetic parameters is
protocol-dependent.

We define C-rate (in \unit{\hour^{-1}} units) as
\begin{equation}
{\rm C}(t) = \frac{3600\,i_{\rm app}(t)}{\mathcal{Q}_{\rm nom}}\,,
\end{equation}
where the nominal capacity, given in SI units in the above equation ($\unit{\coulomb\per\metre\squared}$), is defined as the areal charge density withdrawn from the prescribed initial state before either electrode reaches its independently specified nominal limit given in Table~\ref{tab:benchmark-current-summary}:
\begin{equation}
\mathcal{Q}_{\rm nom} = \min\left[\mathcal{Q}_{\mrn}(x_{\mrn,0}-x_{\mrn,\min})\,,
\mathcal{Q}_\mrp\,(x_{\mrp,\max}-x_{\mrp, 0})\right]\,.
\end{equation}
These nominal limits are used only to define the C-rate scale and are not imposed elsewhere.
\begin{table}[htb]
\centering
\small
\begin{tabular}{lll}
\toprule
Quantity & Symbol & Value \\
\midrule
Duration & $t_{\rm end}$ & $6967\,\unit{\second}$ \\
Sampling frequency & $f_s$ & $1\,\unit{\hertz}$ \\
Initial negative stoichiometry & $x_{\mrn,0}$ & $0.9014$ \\
Initial positive stoichiometry & $x_{\mrp,0}$ & $0.2661$ \\
Maximum discharge current & $\max I(t)$ & $56.745\,\unit{\ampere}$ \\
Maximum charge current & $-\min I(t)$ & $13.984\,\unit{\ampere}$ \\
Nominal negative-electrode discharge limit & $x_{\mrn,\min}$ & $0.0279$\\
Nominal positive-electrode discharge limit & $x_{\mrp,\max}$ & $0.9084$\\
Nominal areal discharge capacity & $\mathcal{Q}_{\rm nom}$ & $35.82\,\unit{\ampere\hour\per\metre\squared}$\\
Maximum absolute C-rate & $\max |C(t)|$ & $5\,\unit{\hour^{-1}}$ (i.e. 5C) \\
Absolute charge throughput & $\int_0^{t_{\rm end}} |I(t)|\,\dd t$ & $12.56\,\unit{\ampere\hour}$ \\
Net discharge & $\int_0^{t_{\rm end}} I(t)\,\dd t$ & $9.44\,\unit{\ampere\hour}$ \\
\bottomrule
\end{tabular}
\caption{Summary of the WLTP-like current protocol used for the synthetic DFN benchmark.}
\label{tab:benchmark-current-summary}
\end{table}

\section{Inference formulation}
\label{sec:inference}

In this section, we formulate the inverse problem for estimating the SPMe electrochemical and transport parameters from the benchmark current–voltage data.

\subsection{Inference parameters}
We infer the effective capacity, kinetic and transport parameter groups of the reduced SPMe representation, whose independent characterisation generally requires dedicated electrochemical experiments and additional modelling assumptions. The electrode thickness fractions $\lambda_k$ and relative porosities $\hat\varepsilon_k$ are instead treated as independently known cell-design quantities. These may be obtained from manufacturer or design information, or from independent characterisation where available.
The inferred set contains three functional parameters in addition to the scalar parameters. To obtain a finite-dimensional inference problem, we represent each function using a finite number of coefficients.
For the solids, we introduce a Fourier expansion, truncated at fourth order:
\begin{equation}
    \log_{10} D^\star_k(x_k)
    =
    \sum_{m=1}^{4}
    \left[
        a_{k,m}\cos(2\pi m x_k)
        +
        b_{k,m}\sin(2\pi m x_k)
    \right],
    \qquad k\in\{\mrn,\mrp\}\,.
    \label{eq:fuji9}
\end{equation}
This form is consistent with the normalisation in Eqs.~\eqref{eq:diffusivity_scalings} and~\eqref{eq:mean_diff}, under which the mean of $\log_{10}D_k^\star$ over $x_k\in[0,1]$ vanishes. This represents each solid diffusivity shape using eight coefficients:
\begin{equation}
\vect{\theta}_{D_k} = \left(a_{k,1},\, a_{k,2},\,a_{k,3},\,a_{k,4},\,b_{k,1},\, b_{k,2},\,b_{k,3},\,b_{k,4}\right)\,.
\label{eq:thetaf9}
\end{equation}
For the electrolyte, we adopt the parameterisation:
\begin{equation}
D^\star_\mre (\hat{c}_\mre)= {\rm e}^{-\beta_\mre \left(\hat{c}_\mre -1\right)}\,,
\end{equation}
which satisfies $D_\mre^\star(1)=1$, consistent with the normalisation in Eq.~\eqref{eq:De_normalisation}.
Here, $\hat c_\mre=1$ denotes the prescribed initial concentration, while $\beta_\mre>0$ controls the concentration dependence. The restriction $\beta_\mre>0$ is consistent with experimentally reported electrolyte diffusivities that decrease with increasing concentration, see e.g.~\cite{Valoen2005}.

With this decomposition, we can now write the parameter vector in terms of these scalar parameters as:
\begin{equation}
\vect\theta = \left(
                    \tau_\mrn,\,
                    \mathcal{Q}_\mrn,\,
                    i^{\rm ref}_{\mrn,0},\,
                    \tau_\mrp,\,
                    \mathcal{Q}_\mrp,\,
                    i^{\rm ref}_{\mrp,0},\,
                    \tau_\mre,\,
                    \mathcal{Q}_\mre,\,
                    \beta_\mre,\,
                    t^+,\,
                    \vect\theta_{D_\mrn},\,
                    \vect\theta_{D_\mrp},\,
                    \mathcal{R}_{\rm ohm}
                    \right)\,,
                    \label{eq:THETA}
\end{equation}
with a total of 27 scalar physical model parameters.

\subsection{Artiphy surrogate forward maps}
\label{sec:surrogate-forward}

The repeated SPMe evaluations required for inference are accelerated using surrogate models implemented within the Artiphy framework. For an imposed current protocol $\vect I$ and parameter vector $\vect\theta$, these provide a differentiable approximation to the SPMe forward map, returning the voltage trajectory $\tilde{\vect V}(\vect I;\vect\theta)$ together with relevant solid and electrolyte states. The surrogate evaluation does not alter the SPMe model specification given above.

The surrogate models are trained and validated over parameter domains based on prior distributions, together with a broad range of current profiles. The parameter priors are physically informed, combining literature-based ranges from LiionDB \cite{Wang_2022} with model-specific coverage requirements (see \ref{app:priors} for details). Both the parameter ranges and current profiles considered in this paper lie within the corresponding training and validation regimes. In particular, the current-profile range includes profiles with peaks up to $5$C. Across these regimes, the surrogate model achieves a millivolt-scale forward accuracy. Surrogate accuracy is independently re-assessed after inference against a conventional numerical implementation of the same SPMe in \ref{app:surrogate}.

Once one-off setup costs are excluded, evaluation of the complete 6,968-point WLTP trajectory takes approximately $4~\unit{\milli\second}$ with Artiphy on an NVIDIA RTX 5000 Ada Generation GPU. The corresponding converged numerical SPMe calculation with PyBaMM/IDAKLU takes approximately $2~\unit{\second}$ using one thread of an Intel Core i9-14900K CPU, giving an indicative speed-up of about 500-fold.

\subsection{Likelihood, priors, penalties and posterior distribution}
Let
\begin{equation}
  \mathcal{D}=\{(t_i, I_i, V_i)\}_{i=1}^{N}
\end{equation}
denote the benchmark dataset consisting of current and voltage time series sampled at $1~\unit{\hertz}$, with $N=6968$.
Let $\vect I=(I_1,\ldots,I_N)$ denote the imposed current protocol, and let $\tilde V_i(\vect I;\vect\theta)$ be the voltage prediction obtained from the Artiphy surrogate framework at $t_i$, with parameter vector $\vect\theta$. We adopt an independent and identically distributed (iid) Gaussian model\footnote{
The independent Gaussian error model is adopted here as a pragmatic working likelihood, rather than as a literal stochastic model of the SPMe--DFN discrepancy. The benchmark is deliberately noise-free, and the structured SPMe--DFN discrepancy is left unmodelled so that its effect on the inferred parameters can be examined directly.
Introducing an explicit discrepancy model would add a further modelling layer; unless informed by suitable prior constraints, discrepancy and physical parameter effects can become confounded \cite{Brynjarsdottir2014}.
The resulting residuals may nevertheless contain temporally correlated contributions from the SPMe--DFN mismatch and from the surrogate approximation.
Since these contributions need not share a common covariance structure, we do not impose an explicit correlation model here. As a result, $\sigma$ should be interpreted as an effective voltage-error scale rather than as the standard deviation of measurement noise.
\label{foot:iid}}
for the residuals:
\begin{equation}
r_i(\vect\theta) = V_i- \tilde{V}_i(\vect{I}; \vect\theta) \overset{\rm iid}{\sim} \mathcal{N}(0,\sigma^2)\,.
\end{equation}
Under this error model, the likelihood, i.e. the probability density assigned to the observed voltage series, conditional on the parameter vector $\vect{\theta}$, and the voltage-error standard deviation $\sigma$, is:
\begin{align}
  f(\mathcal{D}\vert\vect\theta,\sigma)
  &=\frac1{(2\pi\sigma^2)^{N/2}}
  \exp\left[-\frac{1}{2\sigma^2}
  \sum_{i=1}^{N}
  \left(V_i-\tilde V_i(\vect{I};\,\vect\theta)\right)^2\right] \\
  &=\frac1{(2\pi\sigma^2)^{N/2}}
  \exp\left[-\frac{N}{2\sigma^2}
  \operatorname{MSE}(\vect\theta)\right]\,,
  \label{eq:likelihood}
\end{align}
where we used the shorthand for the mean-squared error $\operatorname{MSE}(\vect\theta)=\frac{1}{N}\sum_{i=1}^{N}r_i(\vect\theta)^2$.

We construct prior distributions for the parameters informed by the literature compilation LiionDB~\cite{Wang_2022}. Further details are given in \ref{app:priors}. Assuming prior independence between $\vect\theta$ and $\sigma$, the joint prior density is
\begin{equation}
\Pi(\vect\theta,\sigma)
=
f(\vect\theta)f(\sigma)\,,
\end{equation}
where $f(\vect\theta)$ and $f(\sigma)$ are the prior distributions of the parameter vector and the voltage-error scale, respectively.
The voltage-error scale $\sigma>0$ is inferred jointly with $\vect\theta$ and represents the standard deviation of the assumed
Gaussian voltage-error model. Through the likelihood, $\sigma$ determines the scale assigned to voltage residuals relative to the parameter prior, while the normalisation factor in Eq.~\eqref{eq:likelihood} prevents its arbitrary inflation.
For compactness, $\Pi(\vect\theta,\sigma)$ denotes the prior density in the coordinates used for numerical inference, written as a function of the corresponding physical parameters. The underlying parameterisation and the precise definition of the reported maximum a posteriori (MAP) estimate are given in \ref{app:priors}.

In addition to the literature-derived parameter prior, we impose soft
physical-consistency constraints. For instance, the solid stoichiometry must remain within $x_k \in [0,1]$ and normalised electrolyte concentration must satisfy $\hat{c}_\mre>0$.
For the electrolyte term, we introduce the penalty:
\begin{equation}
P_\mre(\vect\theta) = \sum_{i=1}^N\left\{\left(\frac{1}{\lambda_\mrn}\int_0^{\lambda_\mrn}\sqrt{[\hat{c}_{\mre}(\hat x,t_i)]_-}\,d\hat x\right)^2+\left(\frac{1}{\lambda_\mrp}\int_{1-\lambda_\mrp}^1\sqrt{[\hat{c}_{\mre}(\hat x,t_i)]_-}\,d\hat x\,\right)^2\right\}\,,
\label{eq:liquid-penalty}
\end{equation}
where we defined $[z]_-={\max}(-z,0)$
($\sqrt{[\hat c_\mre]_-}$ is equivalently the imaginary part of the principal square root of $\hat c_\mre$).
If $\hat c_\mre$ is non-negative throughout both electrode domains, the two integrands vanish and $P_\mre=0$.
A negative predicted concentration in either electrode domain produces a non-zero contribution and hence a positive penalty.

Similarly, we define the penalty for the solid concentrations:
\begin{equation}
  P_\mrs(\vect\theta)
  =\sum_{i=1}^{N}\sum_{k\in\{n,p\}}
  \left[
    \max\left[-x_{k, \rm surf}(t_i),0\right]^2 + \max\left[x_{k, \rm surf}(t_i)-1,0\right]^2
  \right]\,,
  \label{eq:solid-penalty}
\end{equation}
where the surface stoichiometry is defined as $x_{k, \rm surf}(t_i) = x_k(\scriptr_k, t_i)\vert_{\scriptr_k=1}$.

We encode these physical-consistency conditions through the (unnormalised) effective prior density
\begin{equation}
\Pi_{\rm eff}(\vect{\theta}, \sigma) = \Pi(\vect{\theta}, \sigma) \,{\rm e}^{-\alpha_\mre P_\mre(\vect{\theta}) -\alpha_\mrs P_\mrs(\vect{\theta})}\,,
\label{eq:pieff}
\end{equation}
where $\alpha_\mre$ and $\alpha_\mrs$ are the weights controlling the contribution from the penalty terms. Together, the literature-informed prior and the physical-consistency penalties are not merely numerical conveniences: the present voltage data do not independently constrain all parameters in the inferred family, as we demonstrate later in the local identifiability analysis in \S\ref{sec:local-identifiability}.

The parameter prior is constructed from the parameter ensemble used to train the surrogate (see \ref{app:priors}), and therefore also keeps the inference within the region over which surrogate accuracy has been established. An unconstrained maximum-likelihood optimisation could instead enter regions where the surrogate has not been validated, so we retain the prior rather than pursuing a maximum-likelihood estimate.
As discussed in \S\ref{sec:local-identifiability}, the subsequent Fisher analysis excludes this prior contribution in order to assess separately the information supplied by the voltage data. Marginal summaries of the conditioned prior are given in Table~\ref{tab:prior-scalar-marginals}, and the corresponding pointwise prior bands for the three diffusivity functions are shown in Fig.~\ref{fig:prior-diffusivity-functions}.

Finally, Bayes' theorem gives the posterior density, up to normalisation:
\begin{equation}
\Pi(\vect\theta, \sigma \vert \mathcal{D}) \propto f(\mathcal{D}\vert\vect\theta,\sigma) \Pi_{\rm eff}(\vect{\theta}, \sigma)\,.
\end{equation}
We seek parameter values that maximise the posterior density. The objective function we use for the optimisation is the negative of the log-posterior density per observation, up to an additive constant:
\begin{equation}
\mathcal{L}(\vect\theta,\sigma) = \log\sigma + \frac{1}{2\,\sigma^2}\operatorname{MSE}(\vect\theta)
  +\frac{\alpha_\mre}{N}P_\mre(\vect\theta)
  +\frac{\alpha_\mrs}{N}P_\mrs(\vect\theta)
  -\frac{1}{N}\log\Pi(\vect\theta,\sigma)\,,
  \label{eq:objective}
\end{equation}
where division by $N$ is a constant rescaling for a fixed dataset and therefore does not alter the location of the posterior mode. Any dependence on sampling rate arises from treating the observations as iid, rather than from this rescaling.

In the present study, using the benchmark dataset described in \S\ref{sec:synthetic-benchmark}, we set $\alpha_\mre=10^2$ and $\alpha_\mrs=10^4$. The resulting estimates are reported as $(\vect\theta_{\rm MAP},\sigma_{\rm MAP})$ and referred to as the MAP estimates. For compactness, the formulation is written here in terms of the physical parameters; the numerical parameterisation of the correlated prior and the precise definition of the reported mode are given in \ref{app:priors}.

MAP optimisation was performed using a multistart strategy.
One hundred feasible initial parameter sets were drawn from the prior and used as starting points for separate Adam optimisation runs. The runs were ranked by their final posterior objective values. The best-performing run was then continued from its endpoint with an additional low-learning-rate refinement. The final reported estimate was the refined solution with the lowest posterior objective.

\section{Inference Results}
\label{sec:results}

\begin{figure}[ht]
    \centering
        \includegraphics[width=\linewidth]{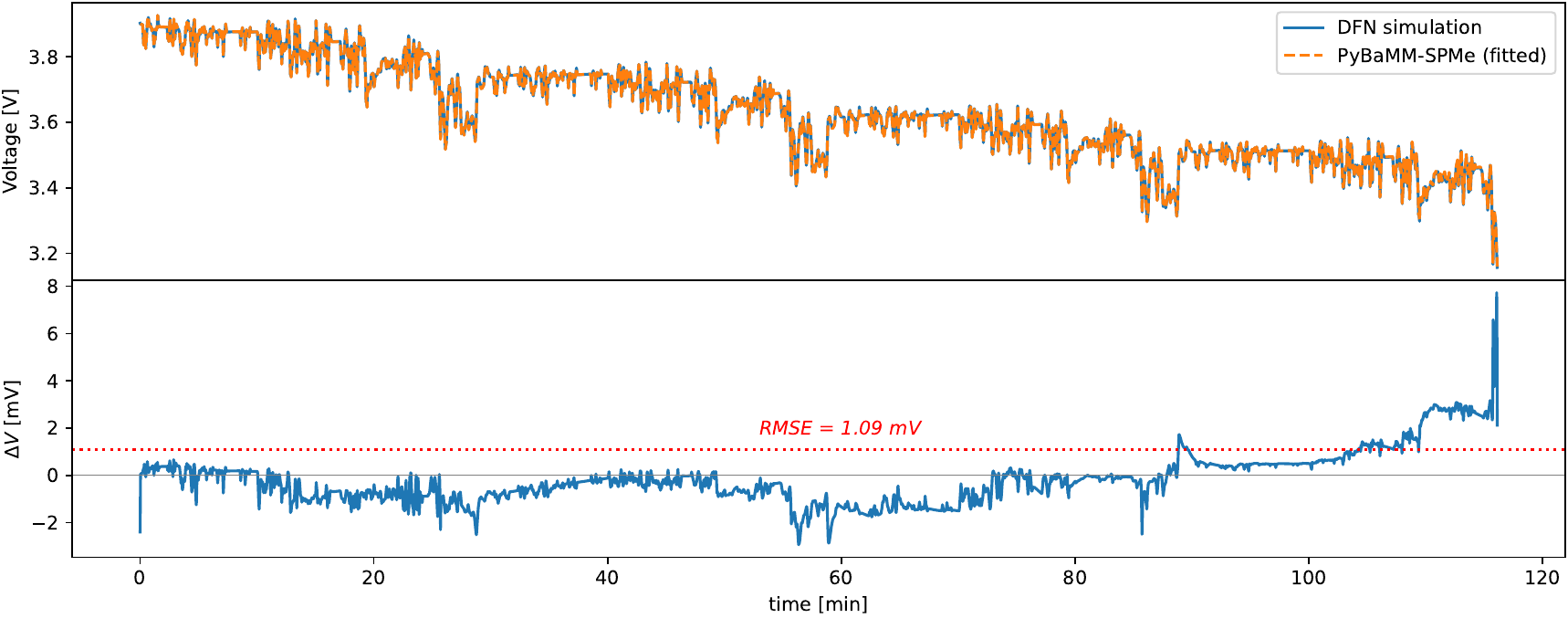}
\caption{
Comparison of the fitted SPMe model prediction using PyBaMM with the original DFN simulation. Top panel shows almost indistinguishable voltage results, while the bottom panel shows the difference $V_{\rm predicted} - V_{\rm DFN}$. The red dotted line denotes the RMSE of $1.09~\unit{\milli\volt}$, with most contribution coming from the last 15 minutes of the protocol.
}
\label{fig:bestfit_pybamm_vs_synth}
\end{figure}

The best refined solution from the multistart optimisation described in \S\ref{sec:inference} yielded a surrogate voltage RMSE of $0.73~\unit{\milli\volt}$ and an effective voltage-error scale of $\sigma_{\rm MAP}=0.73~\unit{\milli\volt}$, reflecting model discrepancy and surrogate approximation error in the noise-free simulation. To verify that the strong fit survives removal of the surrogate approximation, we independently re-evaluated the inferred MAP parameter set using PyBaMM's conventional numerical SPMe solver under isothermal settings.
In Fig.~\ref{fig:bestfit_pybamm_vs_synth}, we show a comparison of the predictions of the fitted SPMe model and the original DFN synthetic data from which the parameters were inferred. Evaluated with this conventional numerical SPMe, the RMSE increases slightly to $1.09~\unit{\milli\volt}$ (represented by the red dotted line in the figure). The largest discrepancies contributing to this value occur near the end of the protocol, consistent with the surrogate's error distribution validated in \ref{app:surrogate}.

\subsection{Parameter estimates}
\label{sec:est_sca}
We now evaluate the inferred values of the parameters. Table~\ref{tab:inferred_scalars} summarises the inferred MAP estimates, the projected DFN benchmark values, and their corresponding relative differences.
Excluding the diffusivity scales and shape parameters, all relative differences remain within $8\%$.
The solid-phase capacities and transference number are all recovered within approximately $2.2\%$ of their projected benchmark values.
\begin{table}[h]
\centering
\begin{tabular}{l c rrr}
\toprule
Parameter & Unit & {Projected benchmark} & {MAP estimate}& {Relative difference [\%]}\\
\midrule
$\tau_\mrn$             & $\unit{\hour}$                            & 2.86                      & 1.46                      & $-49.14^\dagger$    \\
$\mathcal Q_{\mrn}$     & $\unit{\ampere\hour\per\metre\squared}$   & 63.04 	                & 61.65                     & -2.20     \\
$i^{\rm ref}_{\mrn,0}$  & $\unit{\ampere\per\metre\squared}$        & 215.29                    & 220.96                    & +2.63     \\
\midrule
$\tau_\mrp$             & $\unit{\hour}$                            & 3.69                      & 1.85                      & $-49.76^\dagger$    \\
$\mathcal Q_{\mrp}$     & $\unit{\ampere\hour\per\metre\squared}$   & 55.77                     & 55.89                     & +0.21     \\
$i^{\rm ref}_{\mrp,0}$  & $\unit{\ampere\per\metre\squared}$        & 285.70                    & 268.57                    & -6.00     \\
\midrule
$\tau_\mre$             & $\unit{\second}$                          & 144.12                    & 146.99                    & +1.99     \\
$\mathcal Q_{\mre}$     & $\unit{\ampere\hour\per\metre\squared}$   & 1.61                      & 1.68                      & +4.47     \\
$\beta_{\mre}$          & ---                                       & 1.01                      & 0.88                      & -13.45    \\
$t^+$                   & ---                                       & 0.380                     & 0.384                     & +1.08     \\
\midrule
$\mathcal R_{\rm ohm}$  & $\unit{\ohm\metre\squared}$               & {$3.58\times10^{-4}$}    & {$3.31\times10^{-4}$}    & -7.69     \\
\midrule
$\sigma$                & $\unit{\milli\volt}$                      & {---}                     & 0.73                      &           \\
\bottomrule
\end{tabular}
\caption{Inference of model parameters. Entries marked $\dagger$ are solid-phase diffusion timescales and should not be interpreted independently of the corresponding inferred diffusivity-shape parameters. The table omits the 16 solid-diffusivity Fourier coefficients, and the corresponding reconstructed diffusivity functions are presented in \S\ref{sec:est_func}.}
\label{tab:inferred_scalars}
\end{table}

\subsection{Diffusivity reconstruction}
\label{sec:est_func}
\begin{figure}[ht]
    \centering

    \begin{minipage}[t]{1.05\linewidth}
        \centering
        \includegraphics[width=\linewidth]{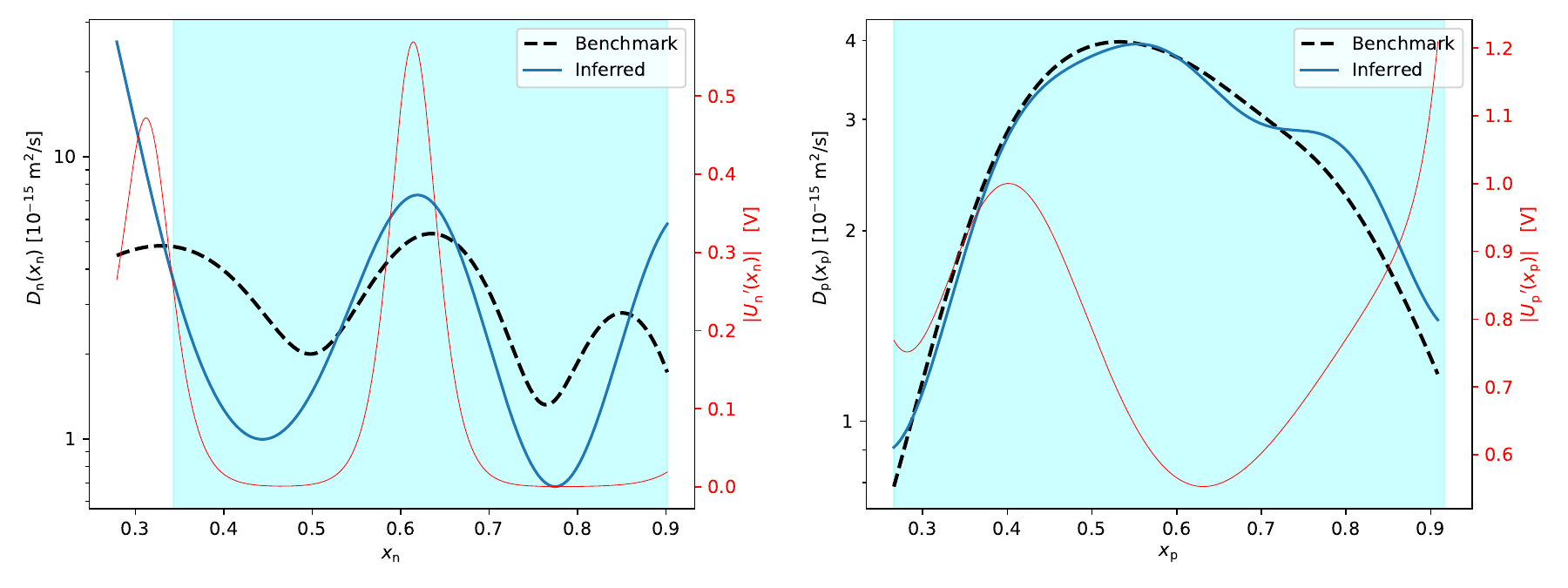}
        \par\smallskip
        \textbf{(a)} Solid-phase diffusivities and OCP gradients.
    \end{minipage}

    \vspace{0.8em}

    \begin{minipage}[t]{0.52\textwidth}
        \centering
        \includegraphics[width=\linewidth]{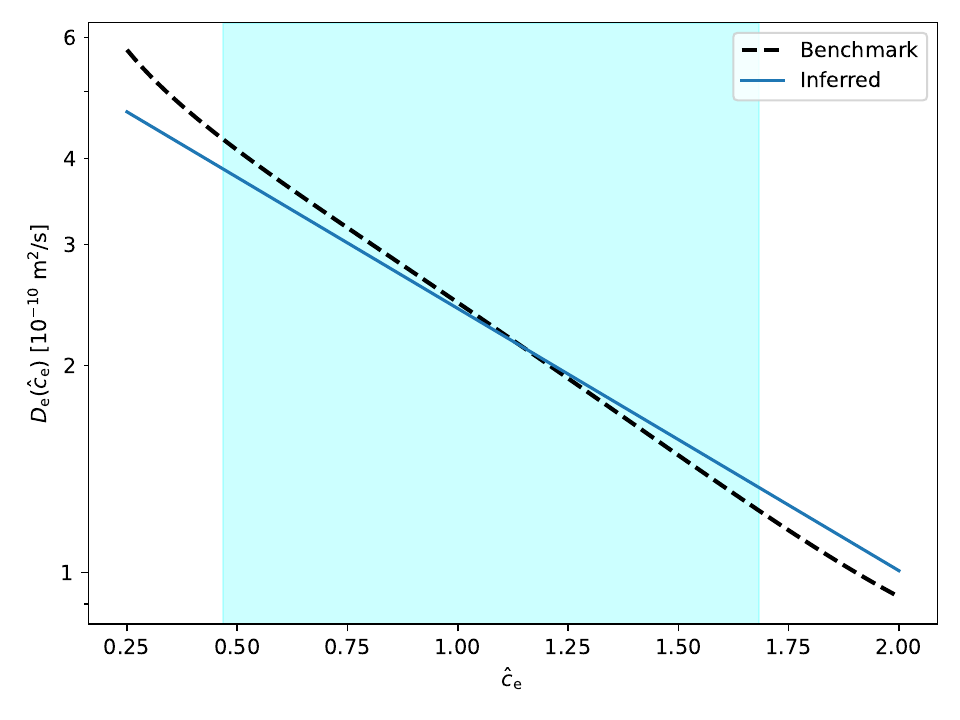}
        \par\smallskip
        \textbf{(b)} Electrolyte diffusivity.
    \end{minipage}

    \caption{
    Inferred non-linear diffusivities from the MAP estimate.
    Solid blue curves show the inferred functions, while dashed black curves show the benchmark functions used in the benchmark simulation.
    Shaded regions indicate the stoichiometry or concentration intervals probed by the drive-cycle trajectory.
    For the solid phases, red curves on the secondary axes show the OCP gradients $|U'_k(x_k)|$, which indicate where stoichiometry perturbations are most visible in the voltage.
    The positive-electrode diffusivity is recovered well over much of the probed range, whereas the negative-electrode diffusivity shows larger recovery errors in flat-OCP regions.
    }
    \label{fig:inferred-transport-functions}
\end{figure}

Figure~\ref{fig:inferred-transport-functions} compares the diffusion functions obtained from the MAP estimate with the DFN benchmark functions used to generate the data. The shaded regions indicate the stoichiometry or concentration intervals visited during the simulated experiment.
Outside these intervals, the inferred curve is not directly constrained by voltage data and should therefore not be interpreted as a data-informed reconstruction of the diffusivity.

The solid diffusion timescales $\tau_k$ should not be interpreted independently of the eight Fourier coefficients defining $D_k^\star(x_k)$ \eqref{eq:fuji9}, since the scale and shape parameters together determine the state-dependent diffusivity. For a given particle radius, $\tau_k$ is inversely proportional to the geometric mean of $D_k$ over the full stoichiometric range, see Eqs.\eqref{eq:diffusivity_scalings}, \eqref{eq:mean_diff}, \eqref{eq:tauk_def}.
Differences in the shape of the diffusivity profile outside the probed range can therefore change the full-range geometric mean. This explains how the substantial differences in $\tau_k$ in Table~\ref{tab:inferred_scalars} can coexist with close agreement over the range probed by the experiment.
We therefore compare the full state-dependent diffusivity profiles in Fig.~\ref{fig:inferred-transport-functions}(a).

Solid-state transport affects the terminal voltage primarily through its effect on the particle-surface stoichiometry entering $U_k$ in Eq.~\eqref{eq:Ueq_unscaled}. Where $U'_k(x_k)$ is small, differences in diffusion dynamics therefore produce only small changes in the OCP contribution to voltage. We indicate these regions by plotting the corresponding OCP gradients, $|U'_k(x_k)|$, on the secondary axes.

The positive-electrode diffusivity is accurately recovered over most of the probed stoichiometry interval. In contrast, the negative-electrode diffusivity exhibits larger discrepancies despite the millivolt-order voltage fit. This behaviour is consistent with the form of the graphite-like negative-electrode's OCP: over extensive portions of the sampled range, the plateaus yield $U_\mrn'(x_\mrn) \approx 0$, rendering the terminal voltage largely insensitive to variations in the negative-electrode concentration profile. Consequently, distinct negative-electrode diffusivity functions can produce virtually identical voltage responses in these flat-OCP regions.

A separate effect arises from the global support of the Fourier representation, discussed in detail in \S\ref{sec:diff_basis}: information acquired over a relatively narrow OCP-sensitive region can constrain coefficients that affect the diffusivity over the full stoichiometric range.

The electrolyte diffusion timescale does not have the same global interpretation as the solid diffusion timescales.
Its reference diffusivity is defined locally at the reference concentration \eqref{eq:Deref_def}, rather than through an average over the concentration range.
The timescale is recovered within $1.99\%$ of its projected benchmark value.
The shape parameter $\beta_\mre$ exhibits a larger relative difference, which may reflect the low sensitivity of the terminal voltage to this parameter, as demonstrated in \S\ref{sec:local-identifiability}.

The complete electrolyte diffusivity is recovered within the concentration range explored by the trajectory, with discrepancies increasing toward the edges of the interval and beyond, consistent with the $13.5\%$ discrepancy in the shape parameter $\beta_\mre$.

\subsection{Voltage tolerance ranges}
\label{sec:voltage-tolerance}

The fitted SPMe accurately reproduces the benchmark voltage even though some inferred constitutive quantities differ substantially from their DFN benchmarks.
We examine how much these functions and the scalar parameters can vary while changing the predicted voltage by no more than a prescribed RMS tolerance.
We retain parameter values for which the root mean square difference between the complete surrogate voltage trajectory and its MAP prediction does not exceed a prescribed tolerance $V_{\rm tol}$.
This difference is evaluated over the same sampling times used for inference.
We use $V_{\rm tol}=1\,\unit{\milli\volt}$, comparable to the independently evaluated SPMe--DFN voltage discrepancy, and $V_{\rm tol}=2\,\unit{\milli\volt}$ to illustrate the effect of relaxing the tolerance.
In each calculation, all parameters other than the scalar or diffusivity block being varied remain fixed at MAP, and candidates are restricted to the corresponding surrogate training domain.
These voltage tolerance ranges characterise parameter variation in the forward map; they are \emph{not} confidence or posterior credible intervals.

\begin{figure}[ht]
\centering
\includegraphics[width=.496\linewidth]{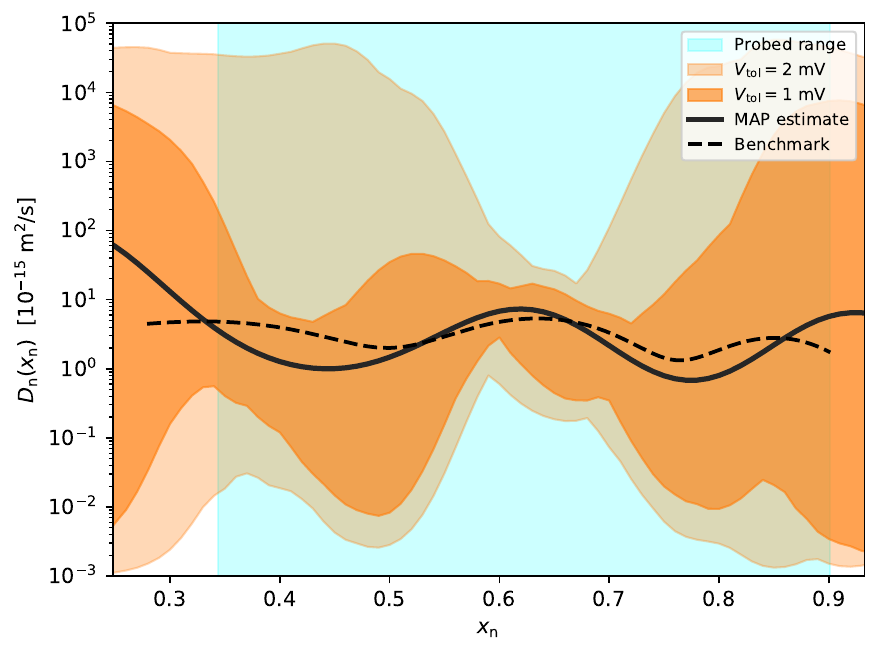}~~
\includegraphics[width=.496\linewidth]{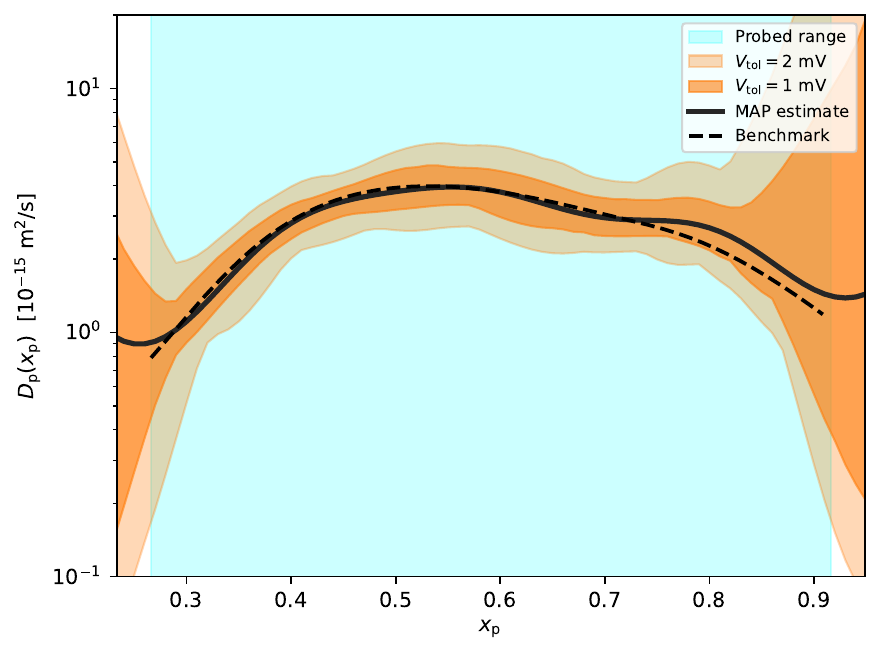}
\caption{{\bf Voltage tolerance envelopes for the solid diffusivities.}
Dark and light orange shading corresponds to
$V_{\rm tol}=1\,\unit{\milli\volt}$ and
$V_{\rm tol}=2\,\unit{\milli\volt}$, respectively.
Here $V_{\rm tol}$ bounds the RMS difference between the complete surrogate voltage trajectory and its MAP prediction.
Each electrode's diffusion timescale and diffusivity-shape coefficients vary jointly, with all remaining parameters fixed at MAP.
Solid black and dashed curves show the MAP and DFN benchmark functions; cyan shading indicates the stoichiometric ranges visited by the MAP trajectory.
The envelopes are pointwise: extrema at different stoichiometries need not arise from the same parameter vector.
These are voltage tolerance envelopes, not statistical uncertainty bands.}
\label{fig:solid_voltage_tolerance}
\end{figure}

For each electrode separately, we vary the diffusion timescale together with the eight Fourier coefficients defining its diffusivity.
At each stoichiometry, constrained numerical optimisation determines the smallest and largest diffusivities compatible with each voltage tolerance.
The resulting envelopes are shown in Fig.~\ref{fig:solid_voltage_tolerance}.
The negative-electrode diffusivity admits particularly large variations in flat-OCP regions, whereas the positive-electrode envelope is narrower over much of the probed range.
Both benchmark functions lie pointwise within the $1\,\unit{\milli\volt}$ envelopes.

For the electrolyte diffusivity, every value of $\beta_\mre$ within the surrogate training domain can be paired with a value of $\tau_\mre$ for which the RMS voltage change remains below $1\,\unit{\milli\volt}$, with all other parameters fixed at MAP.
Thus, the $1\,\unit{\milli\volt}$ criterion does not further restrict the available range of $\beta_\mre$.

\begin{figure}[ht]
\centering
\includegraphics[width=.7\linewidth]{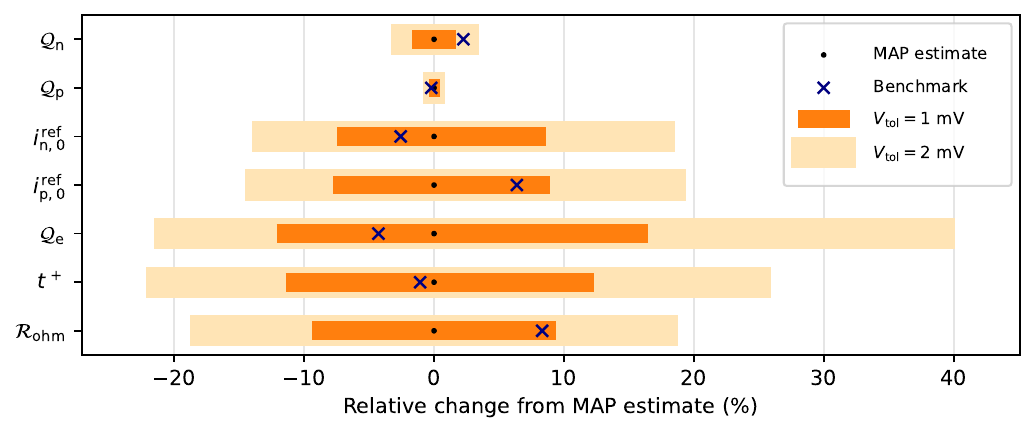}
\caption{{\bf Voltage tolerance ranges for scalar parameters.}
Dark and light orange segments show parameter ranges producing RMS changes of at most $1$ and $2\,\unit{\milli\volt}$ in the complete surrogate voltage prediction relative to MAP.
Each parameter varies individually, with all others fixed at MAP. Black points mark the MAP values, navy blue crosses denote the benchmark values.
The horizontal axis shows percentage changes from MAP.
These are voltage tolerance ranges, not statistical uncertainty intervals.}
\label{fig:scalar_voltage_tolerance}
\end{figure}

We apply the same voltage criterion to the capacities, kinetic scales, transference number and Ohmic resistance, varying each parameter individually with all others fixed at MAP.
The resulting ranges are shown in Fig.~\ref{fig:scalar_voltage_tolerance}.

The electrode capacities admit substantially smaller relative variations than the kinetic and electrolyte-inventory scales.
Numerical bounds and computational details are provided in \ref{app:voltage-tolerance}.

\section{Local sensitivity and practical identifiability}
\label{sec:local-identifiability}

In \S\ref{sec:parameterisation} we eliminated exact parameter redundancies within the SPMe formulation, but this algebraic reduction does not guarantee that the remaining parameter combinations are uniquely identifiable from a finite voltage trace. Then in \S\ref{sec:results}, we demonstrated that the inferred SPMe reproduces the DFN voltage trajectory with millivolt-order accuracy, even though some inferred parameters differ substantially from their projected DFN benchmarks. To investigate this result, we now evaluate the local voltage sensitivity around the MAP estimate to determine which parameter directions are tightly constrained by the benchmark protocol, which remain weakly determined, and which exhibit strong sensitivity correlations. The resulting assessment is local to the MAP estimate, specific to the benchmark protocol, and conditional upon the SPMe model.

The parameter vector $\vect\theta_{\rm MAP}$ \eqref{eq:THETA} contains 27 inferred model parameters. Although the electrode thickness fractions $\lambda_k$ and relative porosities $\hat\varepsilon_k$ were held fixed during inference under the assumption that they are independently known (\S\ref{sec:inference}), we re-introduce them into the sensitivity analysis. This allows us to assess the additional coupling that would arise if these quantities were not known independently. Under the Bruggeman relation adopted in \S\ref{sec:scalings}, isothermal operation at $T = 298.15~\unit{\kelvin}$, and known open-circuit potentials, the local sensitivity analysis therefore contains 31 model parameters.
\subsection{Sensitivity coordinates and voltage Jacobian}
\label{sec:sensi}

The 31 parameters entering the sensitivity analysis have different units, characteristic magnitudes and physical constraints, so sensitivities with respect to the raw parameters are not directly comparable. To compare their local effects on voltage consistently, we transform them into a non-dimensional coordinate vector $\vect\xi\in\mathbb{R}^{31}$, centred on the fitted parameter set. That is, $\vect\xi=0$ denotes the point about which the sensitivity analysis is performed. These transformations place the different parameter directions on roughly comparable local scales.

The parameter transformations are defined as follows. Positive scalar parameters are log-transformed via their ratios to their inferred values, $\log(S / S_{\rm MAP})$. The solid-phase Fourier coefficients $\vect\theta_{D_k}$ \eqref{eq:thetaf9}, which already parameterise log-diffusivity, are centred by subtracting their MAP values, $\vect\theta_{D_k} - \vect\theta_{D_k,\rm MAP}$. The transference number is mapped using a centred logit function, ${\rm logit}(t^+)-{\rm logit}(t^+_{\rm MAP})$, where ${\rm logit}(x) = \log(x / (1-x))$.\footnote{Since $t^+_{\rm MAP}\in(0,1)$, this coordinate is well defined locally. Its use in the sensitivity analysis does not alter the untruncated inference prior described in \ref{app:priors}.}
Finally, to respect the constraint $\lambda_\mrn + \lambda_\mrs + \lambda_\mrp = 1$ on the fractional thicknesses, we apply an additive log-ratio (ALR) transform centred at the benchmark values, $\log(\lambda_k / \lambda_\mrs) - \log(\lambda_{k, \rm MAP} / \lambda_{\mrs,\rm MAP})$, for $k \in \{\mrn, \mrp\}$. For the fixed parameters $\hat\varepsilon_k$ and $\lambda_k$, which were not inferred as part of $\vect\theta_{\rm MAP}$, the subscript ``MAP'' denotes their fixed benchmark values in Table~\ref{tab:benchmark_structurally_id}.

This yields the complete 31-dimensional sensitivity coordinate vector:
\begin{align}
\vect \xi = \Bigg( &
\log\left(\frac{\tau_\mrn}{\tau_{\mrn, \rm MAP}}\right),\,
\log\left(\frac{\mathcal{Q}_\mrn}{\mathcal{Q}_{\mrn, \rm MAP}}\right),\,
\log\left(\frac{i^{\rm ref}_{\mrn,0}}{i^{\rm ref}_{\mrn,0, \rm MAP}}\right),\,
\log\left(\frac{\tau_\mrp}{\tau_{\mrp, \rm MAP}}\right),\,
\log\left(\frac{\mathcal{Q}_\mrp}{\mathcal{Q}_{\mrp, \rm MAP}}\right),\,
\log\left(\frac{i^{\rm ref}_{\mrp,0}}{i^{\rm ref}_{\mrp,0, \rm MAP}}\right),\,
\nonumber\\
&
\log\left(\frac{\tau_\mre}{\tau_{\mre, \rm MAP}}\right),\,
\log\left(\frac{\mathcal{Q}_\mre}{\mathcal{Q}_{\mre, \rm MAP}}\right),\,
\log\left(\frac{\beta_\mre}{\beta_{\mre, \rm MAP}}\right),\,
{\rm logit}(t^+) - {\rm logit}(t^+_{\rm MAP}),\,
\nonumber\\
&
\vect\theta_{D_\mrn} - \vect\theta_{D_\mrn, \rm MAP},\,
\vect\theta_{D_\mrp} - \vect\theta_{D_\mrp, \rm MAP},\,
\log\left(\frac{\mathcal{R}_{\rm ohm}}{\mathcal{R}_{\rm ohm, MAP}}\right),\,
\log\left(\frac{\hat\varepsilon_\mrn}{\hat\varepsilon_{\mrn, \rm MAP}}\right),\,
\log\left(\frac{\hat\varepsilon_\mrp}{\hat\varepsilon_{\mrp, \rm MAP}}\right),\,
\nonumber\\
&
\log\left(\frac{\lambda_\mrn}{\lambda_\mrs}\right) - \log\left(\frac{\lambda_{\mrn, \rm MAP}}{\lambda_{\mrs, \rm MAP}}\right),\,
\log\left(\frac{\lambda_\mrp}{\lambda_\mrs}\right) - \log\left(\frac{\lambda_{\mrp, \rm MAP}}{\lambda_{\mrs, \rm MAP}}\right)
\Bigg)\,.
\label{eq:xidefined}
\end{align}

The local response of the voltage to these coordinates is described by the voltage Jacobian matrix $\matr J$, with elements
\begin{equation}
J_{ij}
=
\left.
\pdv{V(t_i;\vect\xi)}{\xi_j}
\right|_{\vect\xi=0}\,.
\label{eq:voltage_jacobian}
\end{equation}
The $j$--th column of $\matr J$ therefore gives the time-dependent voltage response to a small perturbation of the $j$th parameter coordinate $\xi_j$. More generally, for a sufficiently small displacement $\Delta\vect\xi$, the voltage response is given by
\begin{equation}
\Delta\vect V \approx \matr J\,\Delta\vect\xi\,.
\end{equation}
The following sections use these voltage-response vectors first to assess the magnitude and temporal structure of individual parameter sensitivities, and then to determine whether different parameter perturbations produce distinguishable voltage signatures.

Although the surrogates are differentiable, millivolt-scale forward accuracy does not by itself guarantee accurate derivatives along exceptionally weak sensitivity directions. If the surrogate error varies with the parameters, even a small variation can be comparable to, or larger than, the true voltage response in a weakly sensitive direction. In the weakest individual direction encountered here, a $1\%$ perturbation in $\beta_\mre$ produces an RMS voltage response of only approximately $1.3~\unit{\micro\volt}$ with the conventional numerical SPMe, compared with approximately $1.9~\unit{\micro\volt}$ from the surrogate. The derivative calculation is therefore attempting to resolve voltage responses far below the surrogate's overall forward-accuracy scale. Therefore, throughout this section we evaluate $\matr J$ at the surrogate-inferred MAP point using a conventional numerical SPMe solver.
The Jacobian derivatives are evaluated by central finite differences in the transformed coordinates using $h=5\times10^{-3}$, after confirming stability against a larger step size.

\subsection{Individual voltage sensitivities and temporal signatures}
\label{sec:step_sensitivity}

We first use the voltage Jacobian to examine the local effect of each scalar parameter coordinate individually. For interpretability, we evaluate the linearised voltage response $\Delta\vect V \approx \matr J\,\Delta\vect\xi$ for a set of reference perturbations in the physical parameters. Positive scalar parameters are increased by $1\%$, the transference number by an absolute increment of $0.01$, and either electrode thickness fraction by $0.01$. For a thickness perturbation, the other electrode fraction is held fixed and the separator fraction changes according to $\lambda_\mrs=1-\lambda_\mrn-\lambda_\mrp$; the corresponding displacement therefore generally involves both ALR coordinates in Eq.~\eqref{eq:xidefined}.

\begin{figure}[ht]
    \centering
        \includegraphics[width=0.7\linewidth]{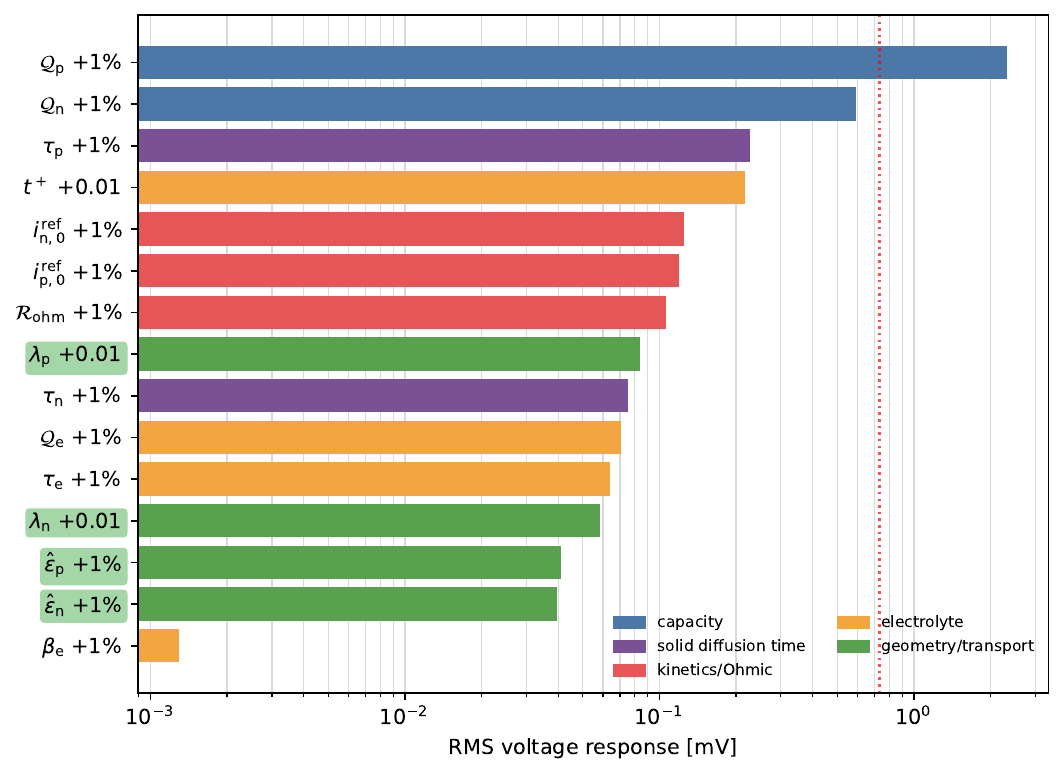}
\caption{
Single-coordinate local voltage sensitivities at the fitted parameter values.
Each bar shows the RMS voltage change produced by the indicated reference perturbation, with all other coordinates fixed.
Positive scalar parameters are perturbed by $1\%$, $t^+$ by $0.01$, and electrode thickness fractions by one percentage point. For a thickness perturbation, the separator fraction changes according to the constraint $\lambda_\mrs=1-\lambda_\mrn-\lambda_\mrp$.
Colours indicate parameter groups.
Parameters whose central values were not inferred in \S\ref{sec:results} are highlighted with green shaded labels.
The dotted line marks the fitted voltage-error scale $\sigma_{\rm MAP}\approx0.73~\unit{\milli\volt}$ and serves only as a scale reference.}
\label{fig:perturbation_bar}
\end{figure}

The resulting RMS voltage responses over the full drive cycle are presented in Fig.~\ref{fig:perturbation_bar}. These provide a simple measure of the magnitude of each individual voltage sensitivity at the fitted parameter set. A large response indicates that the voltage is locally sensitive to that parameter perturbation, whereas a very small response indicates a weakly observed direction. This does not by itself establish identifiability, since two individually strong parameter sensitivities may still produce very similar voltage responses; this is examined in \S\ref{sec:corr_sensitivity}. The dotted vertical line marks the fitted voltage-error scale $\sigma_{\rm MAP}\approx0.73~\unit{\milli\volt}$ as a visual reference only and should not be interpreted as an identifiability threshold.

\begin{figure}[p]
    \centering
        \includegraphics[width=\linewidth]{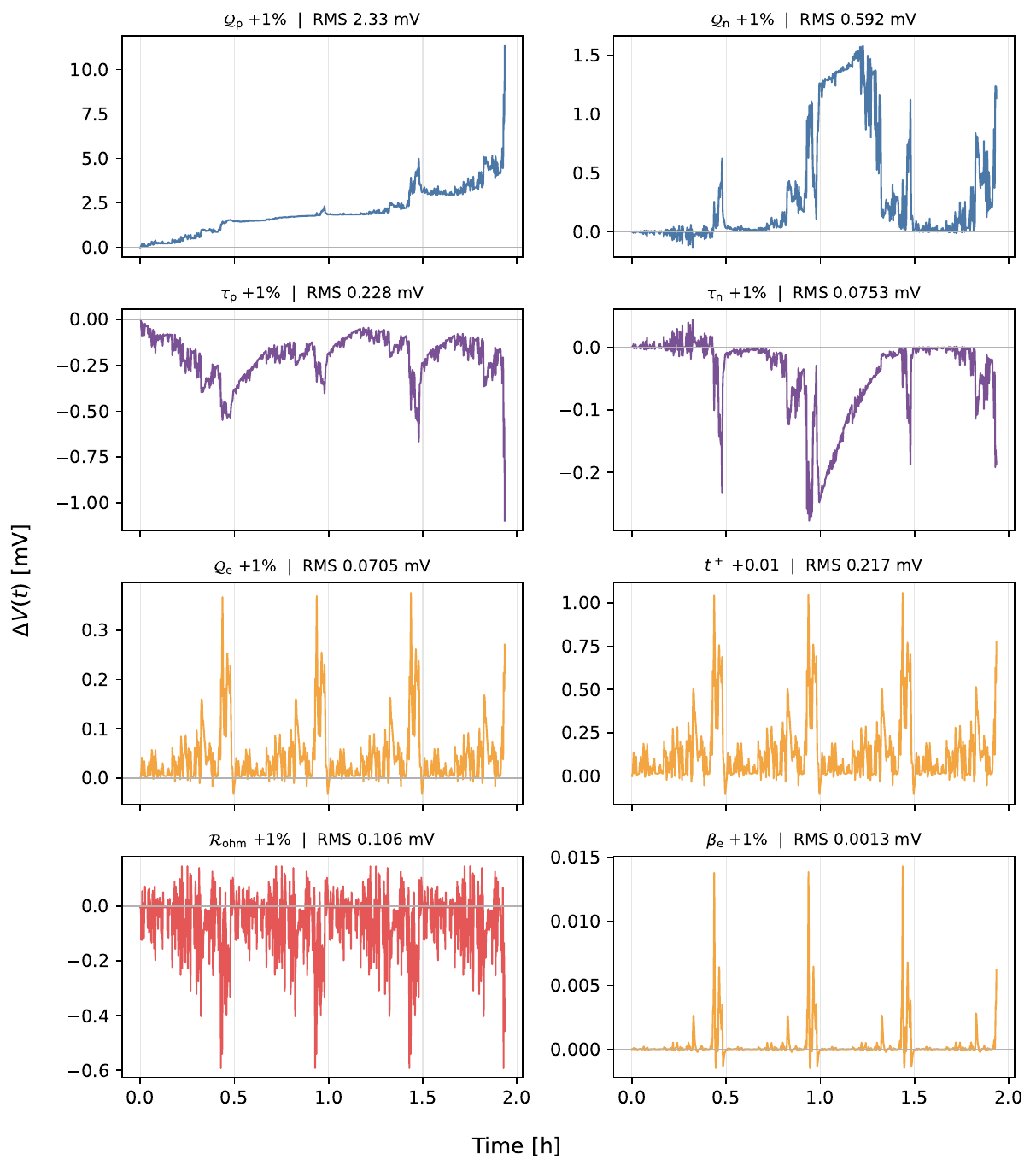}
\caption{
Time-domain voltage responses for representative local reference perturbations at the MAP point.
Each panel shows the linearised voltage change $\Delta V(t)$ for the indicated perturbation, with all other coordinates held fixed.
Perturbation sizes match those in Fig.~\ref{fig:perturbation_bar}.
Panel titles report the RMS voltage response over the full drive cycle.
Note that vertical scales differ between panels; this figure emphasises temporal signatures and qualitative shapes rather than relative magnitudes.
}
\label{fig:perturbation_time}
\end{figure}

The RMS response compresses each sensitivity time series to a single magnitude, whereas its temporal structure can reveal which physical processes generate the voltage response. Figure~\ref{fig:perturbation_time} therefore shows the time-dependent responses $\Delta V(t)$ for representative reference perturbations. Perturbations to the electrode capacities ($\mathcal{Q}_\mrp,\mathcal{Q}_\mrn$) generate cumulative stoichiometric drift. The positive-electrode capacity response is substantially larger, reflecting the greater voltage sensitivity supplied by the positive-electrode OCP along the probed trajectory. Conversely, the response to $\mathcal{Q}_\mrn$ is smaller and more localised in time, developing notable features primarily when the negative-electrode stoichiometry traverses voltage-sensitive OCP regions. The solid-phase diffusion timescales ($\tau_\mrp,\tau_\mrn$) instead modify the diffusion dynamics, producing more transient responses that depend on the recent applied-current history. Consistent with the RMS ranking in Fig.~\ref{fig:perturbation_bar}, the response to $\tau_\mrp$ is substantially larger than that to $\tau_\mrn$.

The remaining responses show that the electrolyte quantities $\mathcal{Q}_\mre$ and $t^+$ produce similar current-driven signatures, with pronounced features coinciding with periods of high current demand where electrolyte concentration gradients are driven most strongly. Their similar temporal profiles already suggest that their effects on voltage may be difficult to distinguish. By contrast, the Ohmic resistance $\mathcal{R}_{\rm ohm}$ follows the applied-current structure directly, as expected for an Ohmic contribution. Finally, the electrolyte shape parameter $\beta_\mre$ displays a temporal signature similar to those of $\mathcal{Q}_\mre$ and $t^+$, but with an extremely small amplitude ($\sim1~\unit{\micro\volt}$ for a $1\%$ perturbation), confirming the weak local voltage sensitivity to this parameter under the benchmark protocol.

Individual sensitivity magnitude alone is therefore insufficient to determine whether two parameters can be distinguished from voltage data: parameter perturbations with similar time-dependent voltage signatures may remain locally confounded even when both produce appreciable responses. We examine these pairwise similarities next.

\subsection{Sensitivity correlations and local parameter confounding}
\label{sec:corr_sensitivity}

While the individual sensitivity analysis in \S\ref{sec:step_sensitivity} quantifies the voltage response to each parameter perturbation, it cannot evaluate pairwise parameter confounding. To quantify the similarity between two sensitivity waveforms, let $\vect J_j$ denote the $j$--th column of the voltage Jacobian \eqref{eq:voltage_jacobian} and define the sensitivity correlation matrix
\begin{equation}
C_{ij} =
\frac{\vect J_i^\top \vect J_j}
{\|\vect J_i\|\,\|\vect J_j\|}\,.
\label{eq:sensicorr}
\end{equation}
Thus, $C_{ij}$ measures whether perturbations in $\xi_i$ and $\xi_j$ change the voltage in the same way over time. Values $C_{ij}\approx 1$ indicate strongly aligned voltage responses, while $C_{ij}\approx-1$ indicates strongly anti-aligned responses; in either case, changes in one parameter can locally mimic or compensate changes in the other. By contrast, $C_{ij}\approx0$ indicates approximately orthogonal voltage signatures and hence greater local distinguishability. The matrix $\matr C$ describes correlations between local voltage sensitivities, rather than posterior correlations between the parameters themselves.

The full $31\times31$ sensitivity correlation matrices evaluated at the fitted MAP estimate and at the projected DFN benchmark point are presented in \ref{app:correlations} (Fig.~\ref{fig:combined_traditional_full_correlation}). The principal conclusions obtained by combining the individual sensitivity magnitudes from \S\ref{sec:step_sensitivity} with these pairwise correlations are summarised in Table~\ref{tab:sensitivity_correlations}.
\begin{table}[ht]
\centering
\small
\renewcommand{\arraystretch}{1.15}
\begin{tabular}{@{}
    p{0.15\linewidth}
    p{0.45\linewidth}
    p{\dimexpr0.40\linewidth-4\tabcolsep\relax}
@{}}
\toprule
Parameter group & Sensitivity and correlation structure & Implication \\
\midrule

$\mathcal Q_\mrn,\mathcal Q_\mrp$
&
Largest scalar sensitivities; distinct from one another, but each is
correlated with its corresponding $\tau_k$.
&
The capacities are distinguishable, although capacity changes can be
partially offset by changes in solid diffusion within each electrode.
\\

$i^{\rm ref}_{\mrn,0}$,
$i^{\rm ref}_{\mrp,0}$,
$\mathcal R_{\rm ohm}$
&
Substantial sensitivities; the kinetic directions are nearly aligned with each other and
nearly anti-aligned with the Ohmic direction.
&
The voltage primarily constrains a combined kinetic--Ohmic loss direction.
\\

$\mathcal Q_\mre,\tau_\mre,t^+$
&
Moderate sensitivities with strongly aligned or anti-aligned voltage
responses.
&
The electrolyte quantities are constrained mainly through correlated
combinations.
\\

$\beta_\mre$
&
Extremely weak response
($\sim 1~\unit{\micro\volt}$ for a $1\%$ perturbation) and correlated with
the electrolyte block.
&
The concentration dependence of $D_\mre$ is very poorly informed.
\\
$\hat\varepsilon_k$ and
$\lambda_k/\lambda_\mrs$
(fixed)
&
Strong geometry--transport coupling. In particular,
$\hat\varepsilon_\mrp$ and $\lambda_\mrp/\lambda_\mrs$ are strongly
correlated, and both couple to electrolyte, kinetic, and Ohmic directions.
The negative-electrode thickness ratio is less strongly entangled.
&
Joint inference introduces substantial additional coupling, especially amongst the geometric and electrolyte transport parameters; see \ref{app:free-structure}.
\\

$\tau_k,\vect\theta_{D_k}$
&
The negative-electrode block is densely correlated; the
positive-electrode block displays greater waveform orthogonality.
&
The negative-electrode diffusivity parameters exhibit stronger local confounding than the positive-electrode parameters.
\\

\bottomrule
\end{tabular}
\caption{Principal local sensitivity magnitudes and pairwise correlation structures under the benchmark drive cycle. Quantities marked as fixed were included in the sensitivity analysis but not inferred in the reported MAP calculation.}
\label{tab:sensitivity_correlations}
\end{table}

The principal sensitivity-correlation structures evaluated at the MAP estimate and at the projected DFN benchmark point are broadly similar (\ref{app:correlations}). In particular, the kinetic--Ohmic confounding and the electrolyte transport block persist at both points, indicating that they are robust local features of the SPMe--protocol combination rather than artefacts of the fitted parameter values alone. The most noticeable dependence on the expansion point occurs within the internal correlation structure of the negative-electrode diffusivity block. We examine the consequences of these diffusivity couplings further in \S\ref{sec:diffusivity}.

\subsection{Local envelopes for the inferred diffusivities}
\label{sec:diffusivity}

The preceding sections considered the magnitude and similarity of the local voltage responses to individual parameter changes.
Here we assess the local sensitivity of the reconstructed diffusivities to voltage perturbations, allowing all inferred parameters to vary jointly near the MAP estimate and their voltage responses to partially compensate. This complements the voltage-tolerance calculation in \S\ref{sec:voltage-tolerance}, which varies one parameter or diffusivity block at a time.

To match the parameter set used in inference, we first remove from $\matr J$ the four columns associated with the fixed geometric and microstructural coordinates: the two relative electrode porosities $\hat\varepsilon_k$ and the two independent thickness-ratio coordinates $\lambda_k$.
Throughout the remainder of this section, $\vect\xi$ and $\matr J$ denote the resulting 27-dimensional coordinate vector and the corresponding reduced voltage Jacobian, respectively.

An inverse-Fisher covariance based on the iid working likelihood reflects the local voltage sensitivities under independent voltage errors. To incorporate prior curvature and temporal residual correlation, we use a local approximation to the fitting objective at the reported MAP estimate. We retain the least-squares weighting with $\sigma=\sigma_{\rm MAP}$, linearise the conventional SPMe voltage response, and retain the local curvature of the prior. In this approximation, perturbations to the observed voltage induce joint changes in the fitted coordinates.
We consider voltage perturbations for which model discrepancy and any measurement noise together are comparable in scale and temporal correlation to the observed residuals.
For these perturbations, we use the covariance estimated from the centred SPMe residuals under a stationarity assumption.
The resulting Gauss--Newton sandwich covariance $\matr C_{\rm GN}$ is defined in \ref{app:local_covariance}.

\begin{figure}[ht!]
    \centering
    \begin{minipage}[t]{\textwidth}
        \centering
        \includegraphics[width=\linewidth]{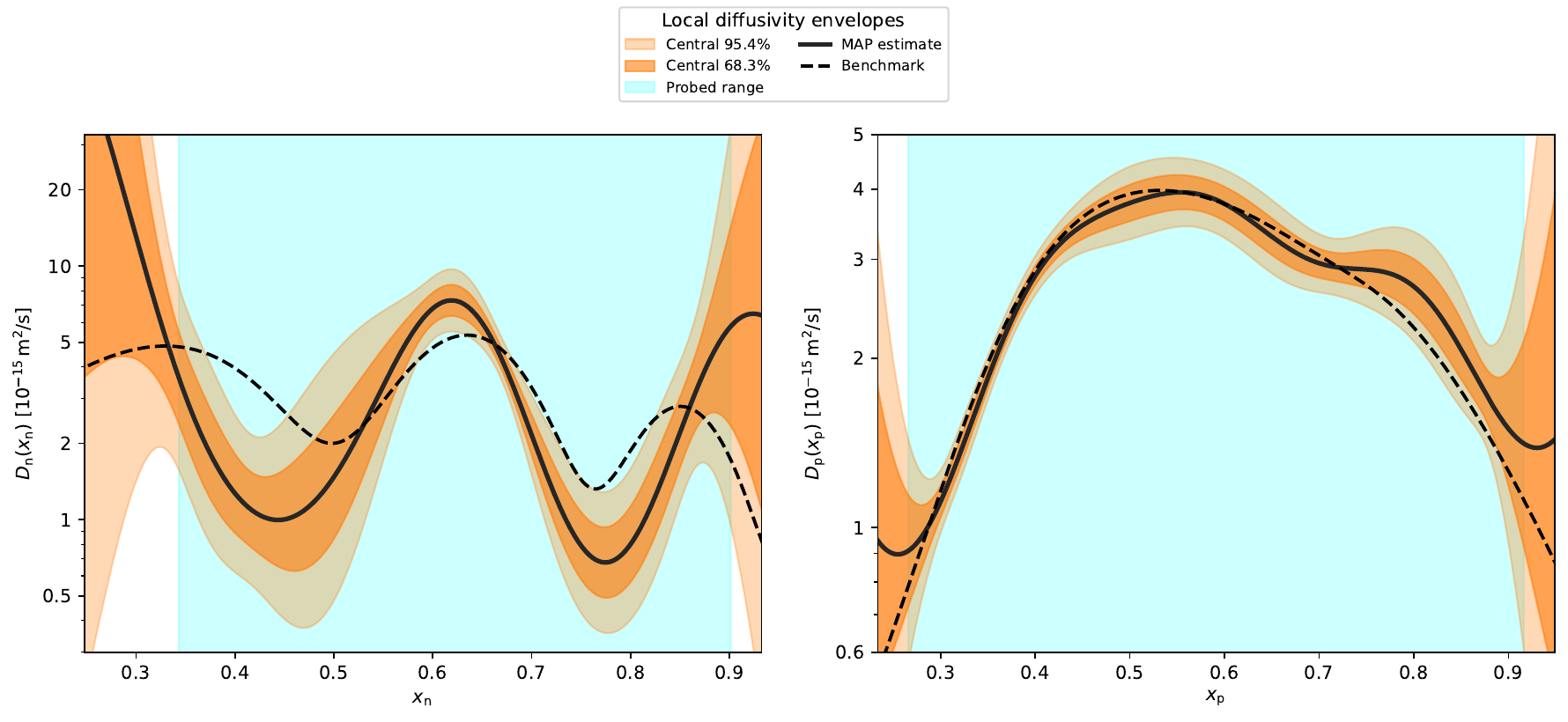}
        \par\smallskip
        \textbf{(a)} Solid-phase diffusivities.
    \end{minipage}

    \vspace{0.8em}

    \begin{minipage}[t]{0.52\textwidth}
        \centering
        \includegraphics[width=\linewidth]{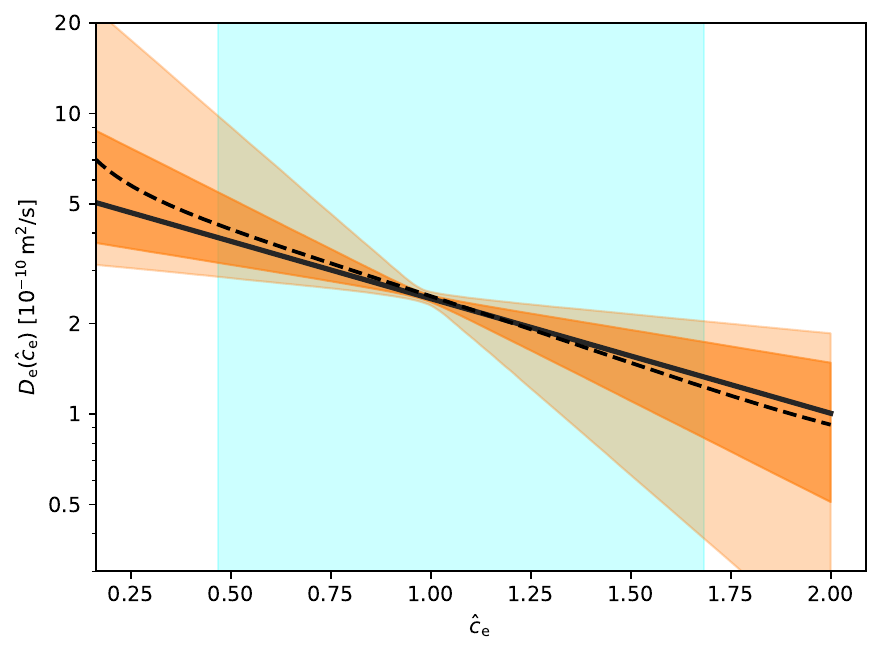}
        \par\smallskip
        \textbf{(b)} Electrolyte diffusivity.
    \end{minipage}

    \caption{
    Local sensitivity of the inferred diffusivities to voltage perturbations.
Solid black curves show the MAP estimates, while dashed black curves show the DFN benchmark functions used in the benchmark simulation.
Shaded cyan regions indicate the stoichiometry (panel a) or concentration (panel b) intervals visited by the drive-cycle trajectory.
Orange regions show the pointwise central $68.3\%$ and $95.4\%$ quantile intervals obtained from MAP-centred Gaussian displacements with covariance $\matr C_{\rm GN}$, propagated through the diffusivity parameterisation.
    }
    \label{fig:local_diffusivity_bands}
\end{figure}

To visualise the resulting variation in function space, we model the MAP-centred parameter displacements as $\vect\xi\sim\mathcal N(\vect 0,\matr C_{\rm GN})$. For a given displacement, we map the transformed coordinates back to the physical parameters and reconstruct the diffusivities \footnote{Reconstructing the diffusivities also conditions on the quantities entering the diffusion-time definitions in Eqs.~\eqref{eq:tauk_def} and \eqref{eq:taue_def}: $R_k$, $L$, $\varepsilon_\mrs$ and $\mathcal B_\mrs$.
In the synthetic benchmark, these are fixed and known exactly.
For experimental data, voltage identifies only the diffusion-time combinations, so these geometric and microstructural quantities would need to be measured independently before the dimensionful diffusivities could be recovered.} as
\begin{align}
D_k(x_k) &= \frac{R_k^2}{\tau_k} D_k^\star(x_k)\,, \qquad \log_{10} D_k^\star(x_k) = \sum_m \left( a_{k,m}\cos(2\pi m x_k) + b_{k,m}\sin(2\pi m x_k) \right)\,,
\qquad k\in\{\mrn,\mrp\}.\nonumber\\
D_\mre(\hat c_\mre) &= \frac{\varepsilon_\mrs L^2}{\mathcal B_\mrs\tau_\mre} \exp\left[-\beta_\mre(\hat c_\mre - 1)\right]\,.
\label{eq:local_diffusivity_mapping}
\end{align}
The spread of the resulting ensemble translates the joint parameter-displacement scale into variations of the reconstructed functions.
The resulting local diffusivity envelopes are shown in Fig.~\ref{fig:local_diffusivity_bands}.

At each fixed stoichiometry or concentration, the plotted regions contain the central $68.3\%$ and $95.4\%$ of the diffusivity distribution. They are therefore pointwise envelopes: the stated fractions apply separately at each abscissa and do not imply that an entire diffusivity curve lies within the corresponding region.

The positive-electrode benchmark lies within the outer envelope over much of the probed stoichiometric range. For the negative electrode, parts of the benchmark fall outside the envelope even at probed stoichiometries. The envelopes describe local variation around the fitted SPMe functions. Bias in those functions can persist because of the restricted parameterisation or SPMe--DFN discrepancy.

The solid-phase envelopes describe variation within the global Fourier representation. We next examine which combinations of the diffusivity coordinates are most visible in voltage and when during the drive cycle that sensitivity arises.

\subsection{Diffusivity parameterisation and localisation of information}
\label{sec:diff_basis}

Each Fourier basis function in the solid diffusivity parameterisation has support over the complete stoichiometric interval. Voltage information acquired while the trajectory passes through a sensitive region can therefore constrain Fourier coefficients that also determine the diffusivity elsewhere.
The envelopes in Fig.~\ref{fig:local_diffusivity_bands} describe how the fitted global functions vary under parameter perturbations around the MAP, rather than showing that the diffusivity is independently resolved at every stoichiometry.

To isolate voltage sensitivity from prior and penalty contributions, we use the Fisher matrix of the iid working likelihood on the 27 inferred coordinates. This gives equal weight to each sampled voltage, consistent with the RMS voltage measure used in \S\ref{sec:voltage-tolerance}. With the error scale fixed at $\sigma_{\rm MAP}$, we define the whitened Jacobian and Fisher information matrix at the MAP point by

\begin{equation}
\matr J^{w}
=
\frac{\matr J}{\sigma_{\rm MAP}},
\qquad
\matr F
=
(\matr J^{w})^\top
\matr J^{w}
=
\frac{\matr J^\top\matr J}
{\sigma_{\rm MAP}^2}\,.
\label{eq:fisher_information}
\end{equation}
To examine how this information is distributed within the solid-diffusivity parameterisation, we consider, for each electrode, the nine coordinates comprising the transformed diffusion-time coordinate and the eight Fourier coefficients of $\log_{10}D_k^\star$.
Assuming that $\matr F$ is numerically nonsingular, we define the effective diffusivity information matrix as
\begin{equation}
\matr{F}^{\rm eff}_{D_k}
=
\left[
    \left(\matr F^{-1}\right)_{D_kD_k}
\right]^{-1}\,.
\label{eq:eff_diff_inf}
\end{equation}
The effective matrix $\matr F^{\rm eff}_{D_k}$ is the Schur complement of the $18\times18$ block of $\matr F$ corresponding to the other inferred coordinates.
For a specified displacement of the nine diffusivity coordinates $\Delta\vect\xi_{D_k}$, the quadratic form defined by $\Delta\vect\xi_{D_k}^\top\matr F^{\rm eff}_{D_k}\Delta\vect\xi_{D_k}$ gives the smallest linearised squared voltage response on the adopted error scale that can be achieved by adjusting the other $18$ inferred coordinates.

For each electrode $k$, let $\lambda_{m}$ and $\vect v_{m}$ denote the eigenvalues and orthonormal eigenvectors of $\matr F_{D_k}^{\rm eff}$, ordered by decreasing eigenvalue:
\begin{equation}
\matr F_{D_k}^{\rm eff}\vect v_m
=
\lambda_m\vect v_m\,.
\end{equation}
Each eigenvector specifies a combination of the nine diffusivity coordinates. Larger eigenvalues identify combinations to which the voltage is more sensitive.

We define the normalised information fraction on the $m$--th mode as
\begin{equation}
f_{m}
=
\frac{\lambda_{m}}{\sum_j\lambda_{j}}.
\end{equation}
These fractions quantify how the local diffusivity information is distributed amongst the diffusivity eigenmodes.

\begin{figure}[ht]
\centering
\includegraphics[width=0.95\textwidth]{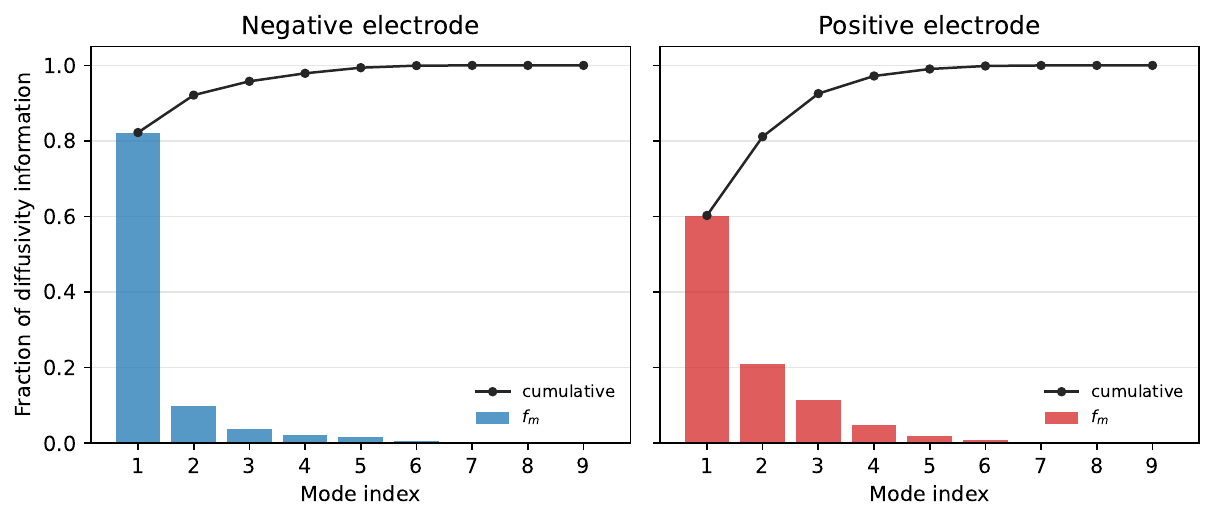}
\caption{
Distribution of the effective local Fisher information within the solid-diffusivity parameterisation.
For each electrode, the bars show the normalised eigenvalue fractions $f_m=\lambda_m/\sum_j\lambda_j$ of the nine-coordinate effective diffusivity information matrix, ordered by decreasing eigenvalue.
The black curves show the corresponding cumulative fractions.
The negative-electrode information is dominated by one mode, whereas the positive-electrode information is distributed more broadly amongst the leading modes.
These fractions refer to the transformed coordinates defined in Eq.~\eqref{eq:xidefined}.
    }
    \label{fig:local_fim_diff_info_block}
\end{figure}

Figure~\ref{fig:local_fim_diff_info_block} shows that the negative-electrode information is strongly concentrated in a single direction: the leading mode accounts for approximately $82\%$ of the total, and the first two modes for approximately $92\%$. The positive-electrode spectrum is less concentrated, with approximately $60\%$ in the leading mode and $81\%$ in the first two. Thus, in both electrodes the effective Fisher information is dominated by a few combinations of the nine diffusivity coordinates, with a more concentrated negative-electrode spectrum and a more distributed positive-electrode one.
The same construction can be used to locate the portions of the trajectory contributing most strongly to a given mode. Expressing the effective Fisher information $\matr F^{\rm eff}_{D_k}$ in terms of its effective Jacobian corresponding only to the diffusivity-related coordinates, we obtain
\begin{equation}
\matr F_{D_k}^{\rm eff} = \left(\matr J_{D_k}^{\rm eff}\right)^\top \matr J_{D_k}^{\rm eff}\,,
\label{eq:effectivefim}
\end{equation}
where $\matr J_{D_k}^{\rm eff}$ contains the residuals over the full drive cycle, obtained by least-squares fitting each of the nine diffusivity-sensitivity columns of $\matr J^w$ using the columns corresponding to other inferred parameters.

\begin{figure}[ht]
    \centering
    \includegraphics[width=0.98\textwidth]{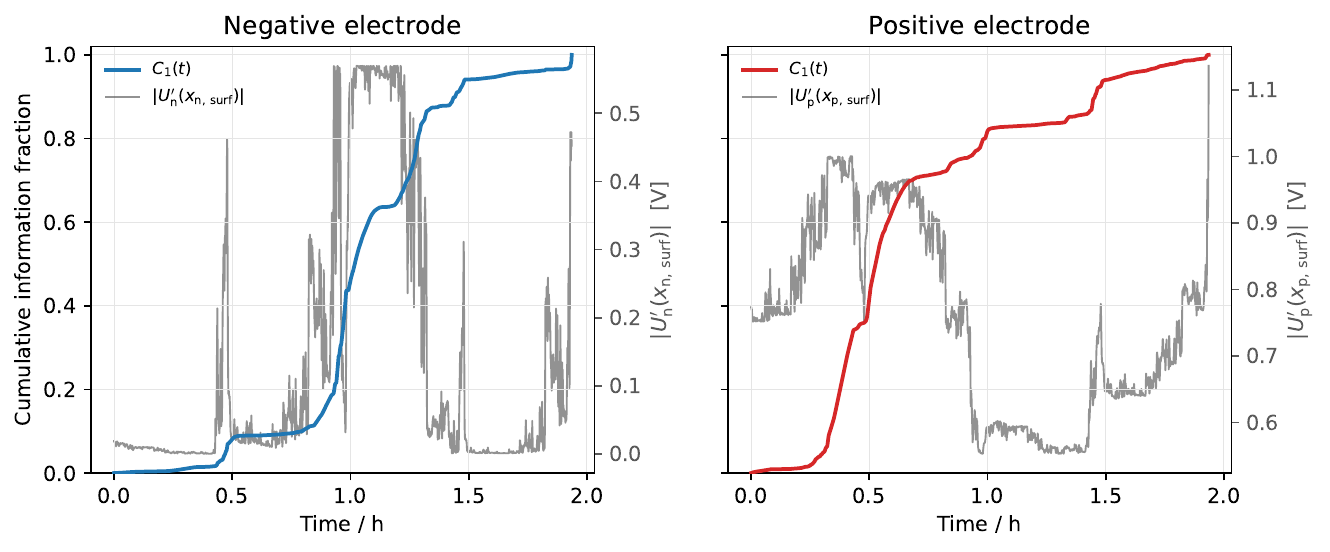}
    \caption{
    Time localisation of the leading diffusivity mode.
    The coloured curves show the cumulative information fraction $C_1(t)$ for the leading diffusivity eigenmode in each electrode.
    Grey curves show the magnitude of the corresponding OCP gradient evaluated along the surface-stoichiometry trajectory, $|U'_k(x_{k,\mathrm{surf}}(t))|$.
    Periods of rapid information accumulation tend to coincide with regions of larger OCP sensitivity, particularly for the negative electrode.
    }
    \label{fig:local_fim_diff_info_time}
\end{figure}

For each electrode, we define the whitened voltage response to the leading diffusivity eigenvector $\vect v_1$ by
\begin{equation}
g_{1}(t_i)
=
\left[
 \matr J_{D_k}^{\rm eff}\vect v_{1}
\right]_i\,.
\end{equation}
Since $\sum_i g_1(t_i)^2=\lambda_1$, we define the cumulative
information fraction as
\begin{equation}
C_{1}(t)
=
\frac{
\displaystyle\sum_{t_i\leq t}g_{1}(t_i)^2
}{
\lambda_1
}\,,
\end{equation}
which is the fraction of the leading mode's full-trajectory squared response contributed by observations up to $t$.
Figure~\ref{fig:local_fim_diff_info_time} compares this accumulation with the
magnitude of the equilibrium-potential gradient evaluated along the
corresponding surface-stoichiometry trajectory.

For the negative electrode, little leading-mode information is accumulated during the early part of the trajectory, whereas a large fraction is acquired as the surface stoichiometry enters the region in which the negative-electrode OCP becomes more sensitive to stoichiometry. The positive-electrode information is accumulated over a broader portion of the drive cycle, with further contributions at later times. In both cases, the association with $|U_k'|$ is expected because the OCP gradient controls how a perturbation in surface stoichiometry changes the equilibrium-potential contribution to the terminal voltage. The correspondence need not be pointwise, however, because the surface-stoichiometry response to a diffusivity perturbation is dynamical and depends on the preceding current history and evolving concentration profile. The construction of the effective FIM in Eq.~\eqref{eq:effectivefim} additionally removes components of the voltage response that can be reproduced by the remaining inferred parameters.

Taken together, Figs.~\ref{fig:local_fim_diff_info_block} and~\ref{fig:local_fim_diff_info_time} show how the effective voltage sensitivity is distributed across diffusivity modes and where along the drive cycle the leading modes are most visible.
The applied drive cycle does not provide uniform, independent information about $D_k(x_k)$ throughout the stoichiometric interval.
Instead, most of the information is concentrated in a small number of directions within the truncated global Fourier parameterisation.
For the leading mode, much of this information is acquired where the trajectory passes through stoichiometric regions with larger OCP gradients, particularly for the negative electrode. Because the same Fourier coefficients determine the diffusivity throughout the stoichiometric interval, these constraints also restrict the reconstructed function elsewhere.

\subsection{Summary}
The local sensitivity and Fisher information analysis reveals clear hierarchies in local voltage sensitivity under the benchmark drive cycle. Terminal voltage is particularly sensitive to the electrode capacities ($\mathcal{Q}_\mrn, \mathcal{Q}_\mrp$) and several solid-phase diffusion directions, particularly for the positive electrode. However, the capacity sensitivities are also correlated with the corresponding solid diffusion timescales. Charge-transfer kinetics ($i^{\rm ref}_{\mrn,0}, i^{\rm ref}_{\mrp,0}$) and Ohmic resistance ($\mathcal{R}_{\rm ohm}$) form a strongly correlated subset. Electrolyte transport parameters ($\mathcal{Q}_\mre, \tau_\mre, t^+$) constitute a tightly entangled correlation block, with the concentration-dependence parameter $\beta_\mre$ remaining exceptionally poorly informed due to both a minute voltage footprint ($\sim 1~\unit{\micro\volt}$ for a $1\%$ perturbation) and strong directional alignment with other electrolyte quantities.
Furthermore, the strong coupling between the currently fixed geometric/microstructural parameters ($\hat\varepsilon_k, \lambda_k/\lambda_\mrs$) and transport directions supports measuring these quantities independently in experimental applications, where possible.
The effect of inferring these quantities jointly is examined in \ref{app:free-structure}.

The envelopes in \S\ref{sec:diffusivity} describe local variation around the fitted SPMe diffusivities. Their widths depend on the linearised voltage response at the MAP, the local effective-prior curvature and the stationary covariance estimated from the centred residuals. Under this construction, parts of the negative-electrode benchmark fall outside the outer envelope within the probed range.
In the Fisher analysis of \S\ref{sec:diff_basis}, the effective diffusivity information is concentrated in a few eigenmodes, more strongly so for the negative electrode than for the positive electrode. These results do not imply that the diffusivity is independently resolved at every stoichiometry.

\section{Conclusions and Discussion}
\label{sec:discussion}
We have developed a surrogate-accelerated framework for inferring a non-redundant set of SPMe parameter groups from dynamic current--voltage data.
To make repeated model evaluations and gradient-based optimisation of the high-dimensional inverse problem computationally feasible, we used a differentiable forward map provided by the Artiphy surrogate framework. The method was tested using voltage data generated by a DFN model rather than by the inference SPMe itself, so that parameter recovery could be examined in the presence of a controlled discrepancy.

For the benchmark drive cycle, the inferred SPMe reproduces the DFN voltage with an RMSE of $1.09\,\unit{\milli\volt}$ when evaluated with a conventional SPMe solver.
The electrode capacities show particularly close agreement with their benchmark values, differing by only $0.21\%$ for the positive electrode and $2.20\%$ for the negative electrode.
Excluding the diffusivity scales and shape parameters, all relative differences remain below $8\%$.
The positive-electrode diffusivity is reconstructed accurately over much of the probed stoichiometric range. The electrolyte diffusivity is also recovered over the sampled concentration range, with discrepancies increasing towards its edges.

The local sensitivity analysis shows that, under the present drive cycle, some parameter changes produce substantial and distinguishable voltage responses, while others have little effect on voltage or can compensate for one another.
The negative-electrode diffusivity differs substantially from the DFN benchmark function despite the excellent voltage fit.
Over extensive portions of the sampled range, the negative-electrode OCP is nearly flat, so differences in diffusion dynamics produce only small changes in the OCP contribution to voltage.

The voltage tolerance calculation demonstrates that substantial negative-electrode diffusivity variations remain compatible with millivolt-scale changes in the predicted voltage.
The dependence of electrolyte diffusivity on concentration remains practically unrestricted within the conditional surrogate domain when its overall scale is allowed to adjust.
A separate effect arises from the global Fourier basis: information acquired during OCP-sensitive portions of the trajectory constrains coefficients that affect the diffusivity over the complete stoichiometric domain. The local diffusivity envelopes describe local variation around the fitted SPMe diffusivities, conditional on the linearised voltage response, the effective-prior curvature and the estimated residual covariance.

The surrogate plays two related roles in this framework. First, it provides a practical avenue for the large number of forward evaluations and gradient-based optimisation steps required by the inverse problem.
Second, its accuracy is established only over the region of parameter space represented in the training distribution, so the prior also helps restrict inference to a domain in which the surrogate has been validated.
At the same time, the sensitivity analysis shows that forward accuracy alone is insufficient to guarantee accurate derivatives in directions whose voltage effect is comparable to or smaller than the surrogate-error scale. We therefore evaluated the final voltage predictions and local sensitivities using conventional numerical SPMe calculations. This hybrid strategy, i.e. surrogate acceleration for inference followed by solver-based verification of physically important results, allows machine-learning acceleration to be exploited without treating the surrogate itself as the physical model.

For experimental inference, however, numerical exactness of the forward solver should not be confused with physical fidelity: even an SPMe (or a DFN model) solved to arbitrary precision remains an approximation to a real cell.
In this setting, surrogate error constitutes an additional controlled approximation, whose acceptability should be judged by its effect on the quantities being inferred.
The present benchmark remains substantially simpler than experimental parameterisation of a real cell,
where additional sources of model discrepancy and uncertainty arise from imperfect model assumptions, OCP uncertainty, thermal effects, and other uncontrolled cell physics.
Future work should therefore consider joint inference from experimental current--voltage data across multiple complementary protocols, with shared intrinsic cell parameters required to explain all datasets simultaneously and predictive performance assessed on operating conditions not used for fitting.
Where available, additional observables and independently measured cell properties such as electrode thicknesses and porosities can further separate weakly identifiable parameter combinations, with protocol selection guided by sensitivity and identifiability analysis.
In the extended inference, freeing electrode thickness fractions and relative porosities broadened the local Fisher envelope for electrolyte diffusivity without improving the independently evaluated voltage fit (\ref{app:free-structure}).
A more systematic treatment of residual correlation and model discrepancy is therefore an important direction for future work.
Surrogate acceleration can remove much of the computational barrier to physics-based parameter inference, while the parameter resolution supported by the experimental data remains determined by the model structure and the information content of the experimental protocols.

\section*{Acknowledgments}
This work was supported by the Faraday Institution (grant number FIRG095). AEG, JB and JF were supported by a Faraday Institution Industrial Sprint project (grant number FIRG080). JP gratefully acknowledges support from Verkor.

\appendix

\section{Electrolyte parameter coordinates}
\label{app:electrolyte_coordinates}

The use of separator properties in the definitions of $\tau_\mre$ and $\mathcal Q_\mre$ in Eqs.\eqref{eq:taue_def}-\eqref{eq:Qe_def}
fixes a convenient reference scale. Replacing these quantities by corresponding whole-cell measures, such as the total salt inventory and an effective diffusion time, would amount to a nonsingular change of coordinates when the remaining dimensionless ratios are retained, and would therefore leave the model and its redundancy count unchanged. This freedom in the choice of reference scale is distinct from the two-dimensional redundancy of the unrestricted electrolyte parameter set discussed below.

Without the Bruggeman relation, the electrolyte concentration problem is written in terms of eight scalar quantities,
\begin{equation}
\lambda_\mrn\,,\quad\lambda_\mrp\,,\quad
\hat\varepsilon_\mrn\,,\quad\hat\varepsilon_\mrp\,,\quad
\hat{\mathcal B}_\mrn\,,\quad\hat{\mathcal B}_\mrp\,,\quad
\tau_\mre\,,\quad\frac{\mathcal Q_\mre}{1-t^+}\,,
\end{equation}
but depends on them only through six combinations. This reduction is obscured in the single global coordinate because $\lambda_\mrn$ and $\lambda_\mrp$ enter through the region boundaries and associated source profiles.
Mapping each region separately onto a unit interval \cite{Northrop2011}, we introduce local coordinates $y_i=(x-x_{i,\rm left})/L_i\in[0,1]$, where $x_{i,\rm left}$ is the left boundary of region $i$. The six regional groups are then
\begin{subequations}
\begin{align}
q_{\mre,i} &= \frac{\mathcal Q_\mre\lambda_i\hat\varepsilon_i}{1-t^+}
=\frac{F c_{\mre,0}L_i\varepsilon_i}{1-t^+}\,,\\
\tau_{\mre,i} &= \frac{\tau_\mre\lambda_i^2\hat\varepsilon_i}{\hat{\mathcal B}_i}
=\frac{\varepsilon_iL_i^2}{\mathcal B_iD_{\mre,\rm ref}}\,,\qquad i\in\{\mrn,\mrs,\mrp\}\,.
\end{align}
\label{eq:regional_electrolyte_groups}
\end{subequations}
These quantities provide a parameterisation equivalent to the corresponding regional groups of Jobman et al. \cite{Jobman2015}, with $\lambda_\mrs=1-\lambda_\mrn-\lambda_\mrp$ and $\hat\varepsilon_\mrs=\hat{\mathcal B}_\mrs=1$.
With the rescaled anion flux
\begin{equation}
\tilde{\mathcal{F}}_- = \frac{F\,\mathcal F_-}{1-t^+}\,,
\end{equation}
the concentration equation within each region becomes
\begin{equation}
q_{\mre,i}\frac{\partial\hat c_\mre}{\partial t}+\frac{\partial \tilde{\mathcal{F}}_-}{\partial y_i}=0\,,\qquad
\tilde{\mathcal{F}}_-=-\frac{q_{\mre,i}}{\tau_{\mre,i}}D_\mre^\star(\hat c_\mre)\frac{\partial\hat c_\mre}{\partial y_i}-i_\mre\,.
\label{eq:regional_electrolyte_dynamics}
\end{equation}
The current profiles are $i_{\rm app}y_\mrn$, $i_{\rm app}$ and $i_{\rm app}(1-y_\mrp)$ in the negative electrode, separator and positive electrode, respectively. Concentration and $\tilde{\mathcal{F}}_- $ are continuous at interfaces, $\tilde{\mathcal{F}}_- =0$ at the current collectors, and $\hat c_\mre=1$ initially. Thus the regional problem contains only the six scalar groups ($q_{\mre, i}$, $\tau_{\mre, i}$) in Eq.~\eqref{eq:regional_electrolyte_groups} and the normalised function $D_\mre^\star$.

Imposing $\hat{\mathcal B}_i=\hat\varepsilon_i^{3/2}$ supplies two relations amongst the original eight quantities. The resulting six coordinates in Eq.~\eqref{eq:electrolyte_lumped} can be recovered uniquely from the positive regional groups \eqref{eq:regional_electrolyte_groups}. To show this, we first observe that the combination
\begin{equation}
q_{\mre,i}\tau_{\mre,i}^2 =\frac{\mathcal{Q}_\mre\,\tau_\mre^2}{(1-t^+)}\,\lambda_i^5\,\,\left(\frac{\hat{\varepsilon}_i^3}{\hat{\mathcal{B}}_i^2}\right)
\end{equation}
becomes independent of porosity under the Bruggeman relation. Then we can write each fractional thickness as
\begin{equation}
\lambda_i = \frac{\left(q_{\mre,i}\tau_{\mre,i}^2\right)^{1/5}}{\displaystyle\sum_{j\in \{\mrn,\mrs,\mrp\}}\left(q_{\mre,j}\tau_{\mre,j}^2\right)^{1/5}}\,,
\end{equation}
where the common factor $\left[\mathcal Q_\mre\tau_\mre^2/(1-t^+)\right]^{1/5}$ cancels upon normalising the three thickness fractions to sum to one. The remaining quantities follow directly:
\begin{equation}
\hat\varepsilon_i = \frac{q_{\mre,i}\lambda_\mrs}{q_{\mre,\mrs}\lambda_i}\,,
\qquad
\tau_\mre = \frac{\tau_{\mre,\mrs}}{\lambda_\mrs^2}\,,
\qquad
\frac{\mathcal{Q}_\mre}{1-t^+}=\frac{q_{\mre,\mrs}}{\lambda_\mrs}\,.
\label{eq:regional_electrolyte_inverse}
\end{equation}
This establishes the one-to-one correspondence claimed in \S\ref{sec:parameter_groups_rescaled}; it does not establish identifiability of these groups from terminal-voltage data.

\section{Prior distributions and latent parameterisation}
\label{app:priors}

The parameter distributions used in the inference delimit the regions over which the surrogate forward maps were validated, while assigning greater weight to physically plausible SPMe parameterisations. The underlying parameter ensembles combine ranges informed in part by the literature compilation LiionDB~\cite{Wang_2022}, with model-specific coverage requirements. The resulting priors should therefore not be interpreted as an empirical joint distribution for a population of cells.

The dependent coordinates are separated into mutually independent Gaussian copula blocks for the negative electrode, positive electrode and electrolyte. For block $b\in\{\mrn,\mrp,\mre\}$, let $F_{b,j}$ denote the fitted marginal cumulative distribution function of its $j$th construction coordinate. The Gaussian-copula transform is
\begin{equation}
\vect z_b\sim\mathcal N(\vect 0,\matr R_b),
\qquad
\psi_{b,j}=F_{b,j}^{-1}\!\left[\Phi(z_{b,j})\right]\,,
\label{eq:latent-prior}
\end{equation}
where $\Phi$ is the standard-normal cumulative distribution function,
$\matr R_b$ is the within-block copula correlation matrix, and $\psi_{b,j}$ is
the corresponding construction coordinate. These marginal maps, together
with the independent scalar marginal maps, define the deterministic mapping
\begin{equation}
(\vect\theta,\sigma)=\mathcal T(\vect z)\,,
\end{equation}
into the manuscript parameters in Eq.~\eqref{eq:THETA}, where $\vect z$ collects the latent coordinates. The copula marginals are truncated to the ranges represented in the validity ensembles. A fourth set contains the independent kinetic, Ohmic, transference-number and voltage-error-scale coordinates, including $\sigma\sim\operatorname{HalfNormal}(5~\unit{\milli\volt})$.

In the reported inference, the solid blocks are conditioned on the prescribed initial stoichiometries and the electrolyte block is conditioned on the prescribed $\lambda_k$, $\hat\varepsilon_k$ and $\hat{\mathcal B}_k=\hat\varepsilon_k^{3/2}$. Holding these entries fixed in the correlated blocks gives the corresponding conditional copula prior for the remaining coordinates. Hereafter, $\vect z$ denotes only those latent coordinates that remain free, with the prescribed values held fixed within $\mathcal T$. Accordingly, $\Pi(\vect\theta,\sigma)$ in the main text is the prior density with respect to these free latent coordinates. The normal prior for $t^+$ is intentionally not truncated: it assigns $0.906\%$ probability below zero and negligible probability above one. This small negative tail
allows $t^+$ to act as an effective parameter when the reduced model does not explicitly represent effects such as ionic aggregation, which can produce apparent negative transference numbers \cite{Richardson2018}.

The objective $\mathcal L$ in Eq.~\eqref{eq:objective} is evaluated through
$\mathcal T$. The reported estimate is therefore
\begin{equation}
\begin{aligned}
\widehat{\vect z}
&=\underset{\vect z}{\operatorname{arg\,min}}\;
\mathcal L\!\left(\mathcal T(\vect z)\right),\\
(\vect\theta_{\rm MAP},\sigma_{\rm MAP})
&=\mathcal T(\widehat{\vect z}).
\end{aligned}
\label{eq:latent-map}
\end{equation}
The reported MAP is thus the mode in the coordinates actually optimised,
mapped into the physical parameters. No change-of-variables Jacobian is needed
in Eq.~\eqref{eq:latent-map} because the optimisation is performed in
$\vect z$. Writing the posterior density directly in the physical parameters
would introduce that Jacobian and can give a different mode.

\begin{table}[ht]
\centering
\begin{tabular}{lrrrrr}
& \multicolumn{3}{c}{Conditioned prior} &
\multicolumn{2}{c}{Reference points}\\
\cmidrule(lr){2-4}\cmidrule(lr){5-6}
Parameter & 5th & Median & 95th & Benchmark & MAP\\
\midrule
$\tau_\mrn$ [\unit{\hour}]
& 0.0513 & 1.89 & 15.4 & 2.86 & 1.46\\
$\mathcal Q_\mrn$ [\unit{\ampere\hour\per\metre\squared}]
& 10.7 & 28.2 & 91.7 & 63.0 & 61.6\\
$i^{\rm ref}_{\mrn,0}$ [\unit{\ampere\per\metre\squared}]
& 1.46 & 54.6 & $2000$ & 215 & 221\\
\midrule
$\tau_\mrp$ [\unit{\hour}]
& 0.221 & 4.54 & 29.1 & 3.69 & 1.85\\
$\mathcal Q_\mrp$ [\unit{\ampere\hour\per\metre\squared}]
& 30.1 & 51.9 & 95.5 & 55.8 & 55.9\\
$i^{\rm ref}_{\mrp,0}$ [\unit{\ampere\per\metre\squared}]
& 0.998 & 38.3 & $1490$ & 286 & 269\\
\midrule
$\tau_\mre$ [\unit{\second}]
& 33.6 & 178 & 934 & 144 & 147\\
$\mathcal Q_\mre$ [\unit{\ampere\hour\per\metre\squared}]
& 0.861 & 1.95 & 4.27 & 1.61 & 1.68\\
$\beta_\mre$
& 0.139 & 0.738 & 2.19 & 1.01 & 0.878\\
$t^+$
& 0.0947 & 0.314 & 0.531 & 0.380 & 0.384\\
$\mathcal R_{\rm ohm}$ [\unit{\ohm\metre\squared}]
& $1.30\times10^{-4}$ & $5.01\times10^{-4}$ & $2.15\times10^{-3}$
& $3.58\times10^{-4}$ & $3.31\times10^{-4}$\\
$\sigma$ [\unit{\milli\volt}]
& 0.308 & 3.38 & 9.82 & \text{--} & 0.733\\
\bottomrule
\end{tabular}
\caption{Marginal summaries of the conditioned prior in manuscript coordinates. The 5th and 95th percentiles delimit the central 90\% prior interval. The benchmark and MAP columns show their locations within the prior; no benchmark value exists for $\sigma$. These one-dimensional summaries do not imply prior independence.}
\label{tab:prior-scalar-marginals}

\end{table}

All available projected benchmark values and all reported MAP values lie within their respective central 90\% prior intervals. The particularly broad kinetic and solid-diffusion-time marginals reflect the intended coverage of the validity ensembles rather than information supplied by the voltage data.

\begin{figure}[ht]
\centering
\includegraphics[width=\linewidth]{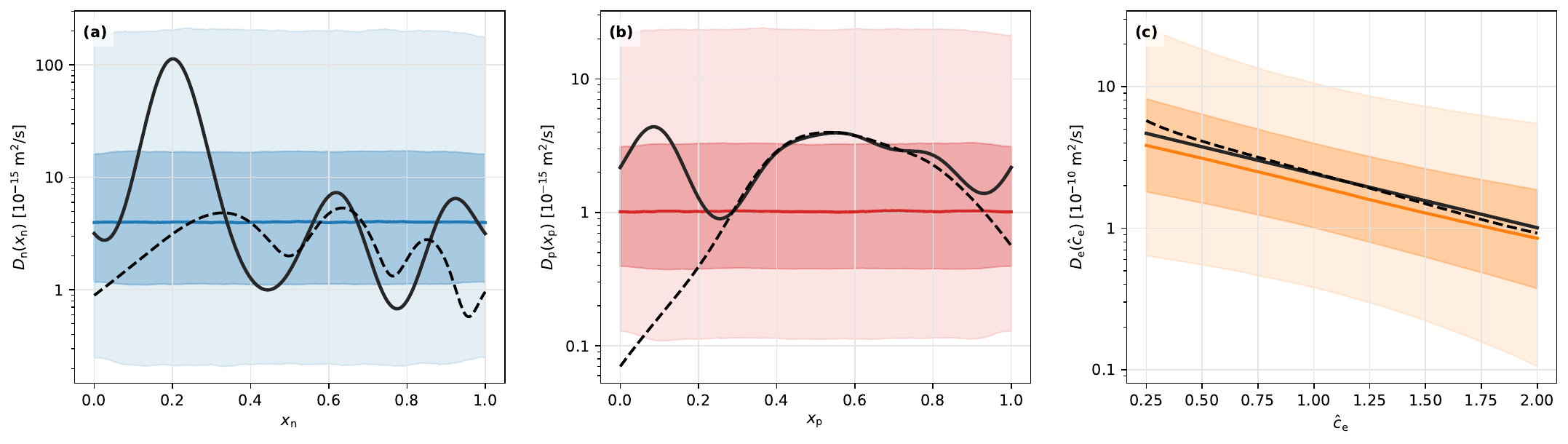}
\caption{Conditioned prior distributions of the dimensional diffusivity functions: (a) negative-electrode $D_\mrn(x_\mrn)$, (b) positive-electrode $D_\mrp(x_\mrp)$, and (c) electrolyte $D_\mre(\hat c_\mre)$. Coloured lines show the pointwise prior medians; darker and lighter shading show the central 50\% and 90\% prior intervals. Solid black lines show the reported MAP functions and dashed black lines the DFN benchmark functions. The electrode bands jointly propagate the diffusion-time and Fourier-shape priors; the electrolyte bands propagate the priors for $\tau_\mre$ and $\beta_\mre$. These are pointwise prior intervals, not likelihood confidence intervals or posterior credible intervals.}
\label{fig:prior-diffusivity-functions}
\end{figure}

\section{Validation of the surrogate}
\label{app:surrogate}

To validate the surrogate independently of the inference calculation, we compare its prediction at the MAP estimate with that of a conventional PyBaMM implementation of the same SPMe. Since the inference returns the non-redundant parameter groups rather than every dimensional model parameter, a dimensional parameter set must first be reconstructed. We retain the independently specified electrode and separator thicknesses, as well as porosities, and set $\varepsilon_k^{\rm act}=1-\varepsilon_k$. The inferred inventories then determine
\begin{equation}
c_{k,\max} = \frac{\mathcal Q_k}{F\,\varepsilon_k^{\rm act}\,L_k}\,,\qquad
c_{\mre,0} = \frac{\mathcal Q_\mre}{F\,L\,\varepsilon_\mrs}\,.
\end{equation}
This reconstruction changes the dimensional concentration scales while preserving the inferred dependence of the constitutive functions on the stoichiometries $x_k=c_k/c_{k,\max}$ and normalised electrolyte concentration $\hat c_\mre=c_\mre/c_{\mre,0}$.

Similarly, for the purposes of the comparison, we fix the particle radii $R_k$ and the diffusion times then determine:
\begin{equation}
D_{k, \rm mean} = \frac{R_k^2}{\tau_k}\,,\qquad
D_{\mre, \rm ref} = \frac{\varepsilon_\mrs L^2}{\mathcal{B}_\mrs \tau_\mre}\,,
\end{equation}
where the transport efficiencies are determined using the Bruggeman relation specified in \S\ref{sec:scalings}.

The inferred Ohmic resistance does not uniquely determine the individual electrolyte and solid-phase conductivities, since these enter the voltage construction only through their combined resistance. For the present reconstruction, we retain the solid-phase conductivities from the projected benchmark parameter set $(\sigma_k^{\rm ref})$ and assign the remaining resistance to the electrolyte:
\begin{equation}
\sigma_\mre(c_{\mre,0})
=
\frac{\frac{L_\mrn}{3\mathcal B_\mrn}
+
\frac{L_\mrs}{\mathcal B_\mrs}
+
\frac{L_\mrp}{3\mathcal B_\mrp}}
{\mathcal R_{\rm ohm}-
\frac{1}{3}
\left(
\frac{L_\mrn}{\sigma_\mrn^{\rm ref}}
+
\frac{L_\mrp}{\sigma_\mrp^{\rm ref}}
\right)}.
\end{equation}
This choice provides one dimensional representative of the inferred lumped resistance and should not be interpreted as separately identifying the electrolyte and solid-phase conductivities.

\begin{figure}[ht]
\centering
\includegraphics[width=\linewidth]{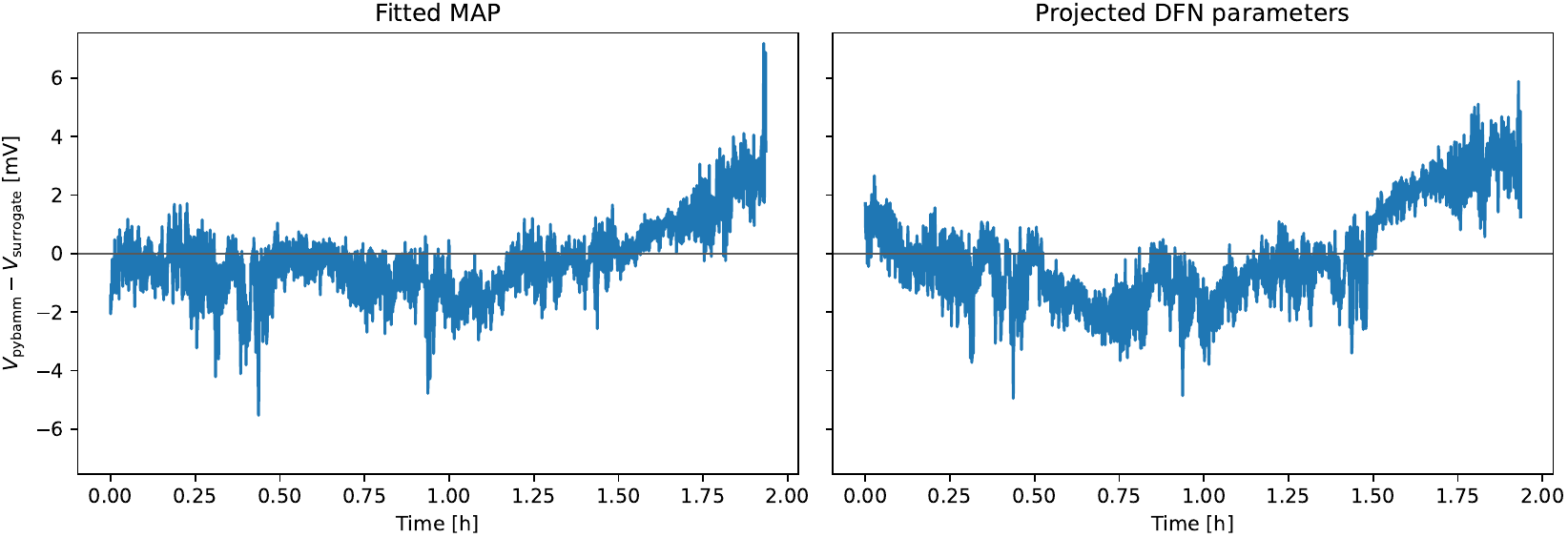}
\caption{Voltage differences between the PyBaMM and surrogate SPMe implementations at the fitted MAP estimate (left) and the projected DFN parameter set (right). The respective RMSE values are $1.25~\unit{\milli\volt}$ and $1.71~\unit{\milli\volt}$, with maximum absolute differences of $7.19~\unit{\milli\volt}$ and $5.89~\unit{\milli\volt}$. The largest discrepancies are localised near the end of the protocol.}
\label{fig:pybamm_minus_surrogate}
\end{figure}

Finally, the inferred reaction-current-density scales determine the corresponding reaction-rate constants:
\begin{equation}
K_k = \frac{i_{k,0}^{\rm ref}\,R_k}{3\,F\,L_k\,\varepsilon^{\rm act}_k}\,.
\end{equation}
A comparison of the surrogate and PyBaMM voltage predictions is shown in Fig.~\ref{fig:pybamm_minus_surrogate}. Across both parameter points, the RMSE remains below $1.8~\unit{\milli\volt}$, with the largest discrepancies localised near the end of the protocol. The residuals contain contributions from both surrogate approximation and differences between the two SPMe implementations. These results support the use of the surrogate as an accurate approximation to the reduced SPMe forward operator over the parameter regime used in the inference.

The Artiphy framework can also be used with alternative OCP functions, once the corresponding initial stoichiometries are specified, and with varying current profiles within the validated input domain, which includes profiles with peaks up to $5$C.

\section{Computation of voltage tolerance ranges}
\label{app:voltage-tolerance}

We apply the same voltage criterion to individual scalar parameters
and to blocks of parameters defining diffusivity functions.
Denoting the varied parameter or block by $\vect\theta_B$, we define
\begin{equation}
\mathcal A_B =
\left\{
\vect\theta_B\in\mathcal S_B :
R(\vect\theta_B)\leq V_{\rm tol}
\right\},
\end{equation}
where $\mathcal S_B$ denotes the domain supported by the surrogate.
Here $R(\vect\theta_B)$ is the RMS change in the complete surrogate
voltage prediction relative to MAP, with all parameters outside
the varied block fixed at MAP.

For an individual scalar parameter $s$, the admissible set
$\mathcal A_s$ directly defines its voltage tolerance intervals.
We determine their endpoints by one-dimensional searches within
$\mathcal S_s$, distinguishing endpoints set by the voltage
tolerance from those set by the surrogate domain boundary.

For a functional diffusivity $D$, we vary its scale and shape parameters
jointly. At each concentration (or equivalently, stoichiometry for solids) $c$, the bounds are
\begin{equation}
D_{\min}(c)
=\min_{\vect\theta_D\in\mathcal A_D}D(c;\vect\theta_D)\,,\qquad
D_{\max}(c)
=\max_{\vect\theta_D\in\mathcal A_D}D(c;\vect\theta_D)\,.
\end{equation}
We use SLSQP to minimise $D$ and $-D$ over
$\mathcal A_D$, obtaining the lower and upper bounds, respectively.
Repeating these optimisations over the concentration grid gives
the pointwise diffusivity envelopes.

Table~\ref{tab:scalar_voltage_tolerance} reports the scalar intervals
for both voltage tolerances.
The first seven rows concern inferred parameters; the final four
concern the thickness fractions and relative porosities prescribed
during inference.

\begin{table}[H]
\centering
\small
\renewcommand{\arraystretch}{1.3}
\setlength{\tabcolsep}{4pt}
\begin{tabular}{l c r c c}
\toprule
Parameter & Unit & MAP estimate
& $1\,\mathrm{mV}$ interval
& $2\,\mathrm{mV}$ interval\\
\midrule
$\mathcal Q_{\mrn}$
& $\unit{\ampere\hour\per\metre\squared}$
& 61.6453
& $[60.6219,\;62.6842]$
& $[59.6022,\;63.7558]$\\

$\mathcal Q_{\mrp}$
& $\unit{\ampere\hour\per\metre\squared}$
& 55.8869
& $[55.6514,\;56.1261]$
& $[55.4200,\;56.3689]$\\

$i^{\rm ref}_{\mrn,0}$
& $\unit{\ampere\per\metre\squared}$
& 220.959
& $[204.471,\;239.935]$
& $[190.010,\;262.017]$\\

$i^{\rm ref}_{\mrp,0}$
& $\unit{\ampere\per\metre\squared}$
& 268.566
& $[247.738,\;292.600]$
& $[229.516,\;320.644]$\\

$\mathcal Q_{\mre}$
& $\unit{\ampere\hour\per\metre\squared}$
& 1.68004
& $[1.47635,\;1.95657]$
& $[1.31798,\;2.35398]$\\

$t^+$
& ---
& 0.384106
& $[0.340220,\;0.431543]$
& $[0.299058,\;0.483606]$\\

$\mathcal R_{\rm ohm}$
& $\unit{\ohm\metre\squared}$
& 0.000330599
& $[0.000299553,\;0.000361645]$
& $[0.000268508,\;0.000392690]$\\
\midrule
$\lambda_{\mrn}$
& ---
& 0.466667
& $[0.300799,\;0.592700^{\dagger}]$
& $[0.153782,\;0.592700^{\dagger}]$\\

$\lambda_{\mrp}$
& ---
& 0.400000
& $[0.277945,\;0.520812]$
& $[0.154623,\;0.525957^{\dagger}]$\\

$\hat\varepsilon_{\mrn}$
& ---
& 0.750000
& $[0.611725,\;1.04411]$
& $[0.527649,\;1.90566^{\dagger}]$\\

$\hat\varepsilon_{\mrp}$
& ---
& 0.800000
& $[0.647686,\;1.07221]$
& $[0.549347,\;1.69960]$\\
\bottomrule
\end{tabular}
\caption{{\bf Scalar voltage tolerance ranges.}
Intervals give the lower and upper parameter values compatible with
$V_{\rm tol}=1$ and $2\,\unit{\milli\volt}$.
Each scalar varies individually, with all other independent parameters
fixed at their MAP or prescribed values.
For the four quantities below the horizontal divider, the MAP column
lists their prescribed values.\\
${}^{\dagger}$ Endpoint set by the surrogate training domain
boundary rather than by the voltage tolerance.}
\label{tab:scalar_voltage_tolerance}
\end{table}

The electrode capacities admit the smallest relative changes, whereas the thickness fractions and relative porosities permit substantially broader ranges.
Several upper endpoints for these prescribed quantities are determined by the surrogate training domain, as indicated by daggers.
These endpoints reflect the available surrogate coverage and should not be interpreted as bounds established by the voltage criterion.

\section{Correlations}
\label{app:correlations}
\begin{figure}[p]
    \centering
    \includegraphics[
        width=\textwidth,
        height=0.76\textheight,
        keepaspectratio
    ]{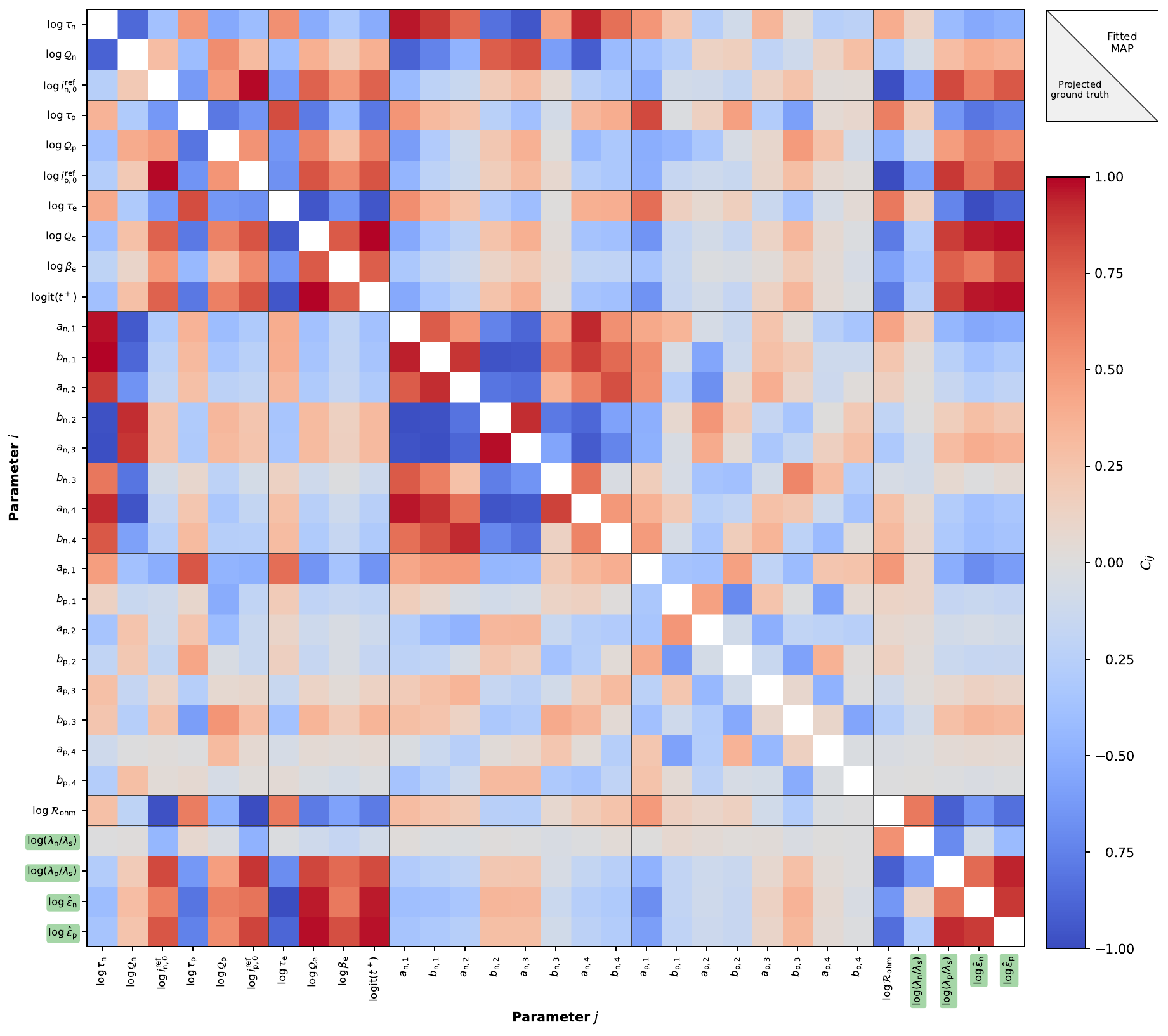}
\caption{\textbf{Normalised local sensitivity correlations for the 31 scaled parameters of the traditional SPMe under the applied current protocol.}
The correlations $C_{ij}$, defined in Eq.~\eqref{eq:sensicorr}, are evaluated at the fitted MAP point (upper-right triangle) and at the DFN benchmark parameters projected onto the SPMe parameterisation (lower-left triangle); both axes use the same parameter ordering. Red and blue indicate aligned and anti-aligned noise-whitened voltage-sensitivity waveforms, respectively, while pale entries indicate approximate orthogonality. Thin lines separate physical parameter groups, and diagonal entries are omitted. The parameters whose central values were not inferred in \S\ref{sec:results} are highlighted in green (shaded label background). These are local sensitivity correlations, not posterior parameter correlations.
    }
\label{fig:combined_traditional_full_correlation}
\end{figure}

Figure~\ref{fig:combined_traditional_full_correlation} compares the normalised
cosine correlations between columns of the noise-whitened voltage Jacobian
associated with the 31 transformed parameter coordinates collected in
$\vect\xi$ \eqref{eq:xidefined}. This provides a two-point
robustness check of the local sensitivity results discussed in
\S\ref{sec:local-identifiability}: in addition to the fitted MAP point, the
Jacobian is evaluated at the benchmark DFN parameters projected onto the SPMe
parameterisation.
Although magnitude information is removed from $C_{ij}$, a direct comparison
of the RMS Jacobian-column norms shows no order-of-magnitude change between
the two parameter points. For 30 of the 31 columns, the norms differ by less
than $9\%$. The exception is $\log\beta_{\rm e}$, whose sensitivity is about
$30\%$ larger at the projected point, although it remains the weakest
individual column by a clear margin.

A separate, like-for-like comparison of each sensitivity trajectory between
the two base points shows that the largest shape changes occur in the
negative-electrode diffusion sensitivities. The cross-point cosine
similarities of the $D_{\rm n}$ Fourier columns range from approximately
$0.887$ to $0.972$, while that of $\log\tau_{\rm n}$ is $0.954$, despite their
RMS norms differing by less than about $9\%$. Most positive-electrode and
remaining scalar sensitivities retain very similar shapes. Thus, the broad
sensitivity hierarchy is stable between these two physically relevant
points, whereas the detailed negative-electrode correlation structure retains
some local base-point dependence.

\clearpage
\section{Gauss--Newton sandwich covariance for the diffusivity envelopes}
\label{app:local_covariance}

We work in the $27$ inferred coordinates used in
\S\ref{sec:diffusivity}, with $\vect\xi=\vect 0$ at the reported MAP
estimate and $\sigma$ fixed at $\sigma_{\rm MAP}$. Let $\vect V$ denote
the observed voltage series and $\tilde{\vect V}(\vect\xi)$ the voltage
predicted by the SPMe at the same observation times for the parameters
corresponding to $\vect\xi$. The Jacobian $\matr J$ is evaluated at
$\vect\xi=\vect 0$.

We approximate the predicted voltage by
$\tilde{\vect V}(\vect 0)+\matr J\vect\xi$. The corresponding approximation
to the summed objective $N\mathcal L$ in Eq.~\eqref{eq:objective},
up to an additive constant, is
\begin{equation}
Q(\vect\xi;\vect V)
=
\frac{\|\vect V-\tilde{\vect V}(\vect 0)-\matr J\vect\xi\|^2}
     {2\sigma_{\rm MAP}^2}
-\log\Pi_{\rm eff}(\vect\xi,\sigma_{\rm MAP})\,,
\end{equation}
where $\Pi_{\rm eff}$ is the prior density, redressed with the penalty terms as introduced in \eqref{eq:pieff}, and is evaluated at the physical parameters corresponding to $\vect\xi$. The curvature of $Q$ at $\vect\xi=\vect 0$
is the Hessian of $Q$, or the Gauss--Newton matrix,
\begin{equation}
\matr H_{\rm GN}
=\left.\nabla_{\vect\xi}^2 Q(\vect\xi;\vect V)
\right|_{\vect\xi=\vect 0}=
\frac{\matr J^\top\matr J}{\sigma_{\rm MAP}^2}
-
\left.
\nabla_{\vect\xi}^2\log\Pi_{\rm eff}(\vect\xi,\sigma_{\rm MAP})
\right|_{\vect\xi=\vect 0}\,.
\end{equation}
For this benchmark, the SPMe states at the reported MAP lie strictly within the physical-consistency bounds, so the penalties contribute zero curvature at that point.

Since a local fit minimises $Q$, its gradient vanishes at the fitted coordinates. For the original voltage $\vect V$ and some hypothetical perturbed voltage $\vect V+\delta\vect V$, we approximate this gradient to first order about the reported MAP.
Subtracting the two zero-gradient conditions gives the change in the fitted coordinates,
\begin{equation}
\delta\widehat{\vect\xi}
=
\frac{1}{\sigma_{\rm MAP}^2}
\matr H_{\rm GN}^{-1}\matr J^\top\delta {\vect V}\,.
\end{equation}

We consider hypothetical voltage perturbations whose scale and temporal correlations are set by the observed DFN-SPMe residual in the MAP fit. We estimate the residual covariance from the observed residuals
$r_i=V_i-\tilde{V}_i(\vect 0)$. Assuming stationarity on the equally
spaced observation grid, we subtract the sample mean
$\bar r=N^{-1}\sum_{i=1}^N r_i$ and use
\begin{equation}
\widehat\gamma(k)
=
\frac{1}{N}\sum_{i=1}^{N-k}(r_i-\bar r)(r_{i+k}-\bar r)\,,
\qquad k=0,\ldots,N-1\,.
\end{equation}
The empirical residual covariance matrix has entries
$(\widehat{\matr\Sigma}_{\rm res})_{ij}=\widehat\gamma(|i-j|)$.
Taking $\widehat{\matr\Sigma}_{\rm res}$ as the covariance of
$\delta{\vect V}$ gives the Gauss--Newton sandwich covariance
\begin{equation}
\matr C_{\rm GN}
=
\frac{1}{\sigma_{\rm MAP}^4}
\matr H_{\rm GN}^{-1}\,
\matr J^\top\,
\widehat{\matr\Sigma}_{\rm res}\,
\matr J\,
\matr H_{\rm GN}^{-1}\,.
\label{eq:local_covariance_gn}
\end{equation}
We use $\matr C_{\rm GN}$ for the MAP-centred Gaussian displacements in
\S\ref{sec:diffusivity}. With
$\widehat{\matr\Sigma}_{\rm res}=\sigma_{\rm MAP}^2\matr I$ and zero
effective-prior curvature, Eq.~\eqref{eq:local_covariance_gn} reduces to
$\matr F^{-1}$, where $\matr F$ is defined in
Eq.~\eqref{eq:fisher_information}.

\section{Effect of inferring electrode thickness fractions and porosities}
\label{app:free-structure}

The inference reported in the main text \S\ref{sec:results} conditions on the electrode thickness fractions and relative porosities. To assess the consequence of treating these quantities as unknown, we repeated the inference after adding $\lambda_\mrn$, $\lambda_\mrp$, $\hat\varepsilon_\mrn$ and $\hat\varepsilon_\mrp$ to the inferred parameter vector. The separator
fraction remains constrained by
$\lambda_\mrs=1-\lambda_\mrn-\lambda_\mrp$, and the transport-efficiency
ratios are updated consistently with the Bruggeman relation,
$\hat{\mathcal B}_k=\hat\varepsilon_k^{3/2}$. All other inferred quantities
and functional parameterisations are retained.

Table~\ref{tab:free-structure-comparison} compares the reported fit with the extended fit. Making the four structural quantities free does not buy a useful improvement in the voltage prediction. The inferred SPMe evolved with PyBaMM reveals an RMSE of $1.208~\unit{\milli\volt}$, a slightly higher value than the previous $1.088~\unit{\milli\volt}$. The original fit is therefore already in a regime
of diminishing returns with respect to terminal-voltage accuracy.

\begin{table}[ht]
\centering
\begin{tabular}{lccc}
\toprule
Quantity
& Projected benchmark
& Inference with $\lambda_k,\hat\varepsilon_k$ fixed
& Inference with $\lambda_k,\hat\varepsilon_k$ free \\
\midrule
RMSE
& --- & 1.088 $\unit{\milli\volt}$ & 1.208 $\unit{\milli\volt}$ \\
\midrule
$\lambda_\mrn$ & 0.467 & 0.467 (fixed) & 0.479 \\
$\lambda_\mrp$ & 0.400 & 0.400 (fixed) & 0.439 \\
$\lambda_\mrs$ & 0.133 & 0.133 (fixed) & 0.082 (derived)\\
$\hat\varepsilon_\mrn$ & 0.750 & 0.750 (fixed) & 0.499 \\
$\hat\varepsilon_\mrp$ & 0.800 & 0.800 (fixed) & 0.348 \\
\midrule
$\tau_\mre$ & 144.12 $\unit{\second}$ & 146.99 $\unit{\second}$ & 118.52 $\unit{\second}$ \\
$\mathcal Q_\mre$
&  1.61  $\unit{\ampere\hour\per\metre\squared}$& 1.68 $\unit{\ampere\hour\per\metre\squared}$& 3.25 $\unit{\ampere\hour\per\metre\squared}$\\
$\beta_\mre$ &1.01 & 0.88 & 0.75 \\
$t^+$ & 0.380 & 0.384 & 0.361 \\
\midrule
\begin{tabular}{@{}l@{}}Median $1\sigma$ width\\ in $\log_{10}D_\mre$\end{tabular}
& --- & 0.088 & 2.543 \\
\bottomrule
\end{tabular}
\caption{Comparison of the reported inference, in which the electrode
thickness fractions and relative porosities are prescribed, with the extended
inference in which they are fitted. The voltage RMSE is evaluated against the
same synthetic data using PyBaMM's SPMe. The final row reports the median full width of the central-$68.3\%$ envelope in $\log_{10}D_{\rm e}$, taken over $\hat c_\mre\in[0.1,2]$.
}
\label{tab:free-structure-comparison}
\end{table}

The most striking difference is in the broadening of the electrolyte diffusivity Fisher envelope, which increases by a factor of 30. In diffusion amplitude terms, this corresponds to a lower-to-upper multiplicative span of approximately $10^{2.543}\simeq349$ in $D_\mre$.
This is caused by the added compensation directions, rather than by the displacement of the fitted point alone.
At the extended fit, conditioning on the four structural coordinates gives a median $1\sigma$ width of $0.038$ decades, whereas allowing them to vary increases it to $2.543$ decades.
The weakest local Fisher direction combines the electrode thickness fractions and porosities with the electrolyte inventory and diffusion timescale. Freeing the four structural quantities therefore gives the electrolyte effects additional directions along which they can be compensated, with the largest consequence for the inferred amplitude of $D_\mre$.

The electrode-capacity and solid-diffusivity conclusions are otherwise essentially unchanged. For this benchmark, the extended calculation therefore supports the conditional analysis used in the main text: independently determining some members of this correlated geometric and electrolyte transport group materially sharpens inference of the remaining members, while the extended fit provides no solver-verified improvement in voltage accuracy.

\bibliographystyle{unsrtnat}
\bibliography{surrbib}

\end{document}